%% file: swampland_sub2.tex
\documentclass[11pt, letter]{article}

\usepackage{stolenstyle}

\usepackage{amsmath,epsf,amssymb,latexsym,amsthm,setspace,bbm,array,pifont,enumerate}

\usepackage[matrix,arrow,frame,import,curve,color]{xy}

\input{basedefs2}

\theoremstyle{definition}

\usepackage{tikz}

\usetikzlibrary[shapes.geometric]
\usetikzlibrary{positioning} 
\usetikzlibrary{calc,intersections,through,backgrounds}

\tikzset{>=stealth}
\tikzset{every picture/.style={very thick}}

\def\cWb{{\overline{\cW}}}

\def\bqb{{{{\qb}}}}
\def\bQ{{{{\boldsymbol{Q}}}}}
\def\bQb{{{{\overline{\boldsymbol{Q}}}}}}

\def\bq{{{{\boldsymbol{q}}}}}

\def\cRb{{{\overline{\cR}}}}

\def\varphib{{\overline{\varphi}}}

\def\WW{{{\mathbb{W}}}}

\def\boldt{\boldsymbol{t}}

\def\tuv{{{\text{uv}}}}
\def\tir{{{\text{ir}}}}

\def\preprint{ ~~ }

\title{Fixed points of (0,2) Landau-Ginzburg renormalization group flows and the chiral algebra}
\author[a] {Marco Bertolini,}
\author[b] { Ilarion V.~Melnikov,}
\author[c] {M.~ Ronen Plesser}
\affiliation[a]{Kavli Institute for the Physics and Mathematics of the Universe (WPI),\\
The University of Tokyo, Kashiwa, Chiba 277-8583, Japan}
\affiliation[b] {Department of Physics and Astronomy \\
James Madison University\\ 
Harrisonburg, VA 22807, USA}
\affiliation[c] {Center for Geometry and Theoretical Physics, Box 90318 \\
Duke University\\
Durham, NC 27708-0318, USA}

\emailAdd{marcomulinex@gmail.com}
\emailAdd{melnikix@jmu.edu}
\emailAdd{plesser@cgtp.duke.edu}

\abstract{We discuss renormalization group flows in two-dimensional quantum field theories with (0,2) supersymmetry.  We focus on theories with UV described by a Landau-Ginzburg Lagrangian and use the chiral algebra to constrain the IR dynamics.  We present examples where the structure of the chiral algebra is incompatible with unitarity of the IR superconformal theory and discuss the implications of this result for programs of classifying (0,2) SCFTs as endpoints of flows from simple Lagrangian theories.}

\begin{document}

\maketitle

\section{Introduction}\label{s:intro}

The renormalization group flow plays a key role in the study of quantum field theory.  In its simplest application it offers an interconnection between two, usually different, theories:  an ultraviolet (UV) theory valid at high energies, and the low energy infrared (IR) theory.  A well-studied situation is where the IR theory is a non-trivial conformal field theory, while the UV theory is Lagrangian and asymptotically free.  In this case we can hope to construct interesting conformal field theories from the simpler UV description.

This is not an abstract hope:  examples of such flows are known, and some key properties of the corresponding conformal field theories are computed explicitly.  For example two-dimensional superconformal field theories (SCFTs) with (2,2) supersymmetry are constructed in this fashion from super-renormalizable two-dimensional gauged linear sigma models~\cite{Witten:1993yc}.  The computable IR quantities go well beyond matching anomalies and global symmetries or even elliptic genera.  For instance, is possible to compute the dependence of the chiral ring of the SCFT on the marginal deformations representable as deformations of couplings in the UV Lagrangian~\cite{Morrison:1994fr}\footnote{For a modern localization perspective the reader can consult~\cite{Closset:2015rna}.  A pedagogical discussion of this, as well as other topics in this paper, is given in~\cite{Melnikov:2019tpl}.} , as well as the sphere partition functions~\cite{Benini:2012ui,Gomis:2012wy}.

While much remains to be understood about the landscape of (2,2) SCFTs, a more mysterious realm beckons: the (0,2) SCFTs.  As already pointed out in~\cite{Witten:1993yc,Distler:1993mk}, the linear sigma model approach generalizes to this case, and in examples explicit computations generalizing the (2,2) results have been performed---see, e.g.~\cite{McOrist:2008ji,Closset:2015ohf}.  More precisely, these results were obtained in (0,2) theories that are smooth deformations of (2,2) theories.

To what extent can these techniques be extended to ``genuine'' (0,2) SCFTs?  Is it possible to classify the  landscape of (0,2) SCFTs obtained from (0,2) linear sigma models?  These questions motivated the present work, and our results suggest that the answers are subtle and will require new tools.

Ultimately, the structure that is responsible for the successes in the analysis of the (2,2) theories and their (0,2) deformations is the chiral algebra~\cite{Witten:1993jg,Witten:2005px}:  the cohomology of the right-moving supercharge $\bQb$.   In the SCFT this cohomology group, which we will denote $H$, is isomorphic to the vector space of right-moving (in our conventions anti-holomorphic) chiral primary operators and includes the left-moving (holomorphic) energy-momentum tensor and any other holomorphic conserved currents.  The operator product, when restricted to cohomology, endows this vector space with the structure of a holomorphic vertex operator algebra.  The topological heterotic rings of~\cite{Adams:2003zy,Adams:2005tc}  are sub-sectors of this vertex operator algebra that generalize the (c,c) and (a,c) rings of (2,2) theories and share some properties with local operators in topological field theory.  

It is believed that under fairly weak assumptions the chiral algebra is a ``renormalization group invariant'' of the theory.
In a wide class of super-renormalizable (0,2) Lagrangian theories the UV $\bQb$-cohomology has a presentation in terms of free fields---a sort of $bc$--$\beta\gamma$ system familiar from the superstring worldsheet~\cite{Fre:1992hp,Witten:1993jg}.\footnote{A modern discussion in the context of (0,2) Landau-Ginzburg models can be found in~\cite{Dedushenko:2015opz}.}   When the theory is defined with a Wilsonian renormalization energy scale $\mu$, there is an argument that the $\bQb$-cohomology is independent of $\mu$ for  $0<\mu < \infty$.\footnote{Early work includes~\cite{Witten:1993jg,Silverstein:1995re}.  A formal argument for the statement in the case of (0,2) Landau-Ginzburg theory is given in~\cite{Dedushenko:2015opz}.}

Consequently, if one assumes a smooth $\mu\to 0$ limit of the chiral algebra structure, it is possible to compute non-trivial properties of the SCFT by doing free-field computations!  Early uses of these structures included the determination of the massless spectrum of heterotic compactifications based on Landau-Ginzburg (LG) orbifolds~\cite{Kachru:1993pg} and constraints on (0,2) RG flows~\cite{Silverstein:1995re}.  Recent work includes the construction of a wide class of so-called hybrid CFTs~\cite{Bertolini:2013xga,Bertolini:2017lcz}.  Some of these results, especially in the context of (2,2) theories, suggest that in many cases the chiral algebra of the SCFT is correctly reproduced by the UV computations.  

A simple example illustrates some of the subtleties of the $\mu\to 0$ limit.  Consider a (2,2) LG theory with a single chiral superfield $\Phi$ with lowest component $\phi$ and a superpotential 
\begin{align}
W = \frac{g_3}{3} \Phi^3 - \frac{g_4}{4} \Phi^4~.
\end{align}
There is a renormalization scheme in which $W$ is not renormalized, and the chiral algebra has an operator $\phi$ and relation $g_3\phi^2 - g_4\phi^3 = 0$.  However, when $g_3$ and $g_4$ are both non-zero, this relation does not reflect the IR physics.  The IR physics depends on the choice of SUSY vacuum, labeled by the expectation value $\la\phi\ra = \phi_\ast$, with either $\phi_\ast = 0$ or $\phi_\ast = g_3/g_4$.  The former yields a massless theory---the $A_1$ N=2 minimal model with (c,c) ring generated by an operator $X$ obeying $X^2 = 0$, while the latter is a massive theory.  

In this example we can resolve the ambiguity:  expanding around either vacuum in terms of a field $X = (\Phi - \phi_\ast)$, and demanding that the R-symmetry acts linearly on $X$, we understand the low energy behavior of the RG flow as either an irrelevant deformation of the $A_1$ SCFT ($\phi_\ast = 0$) or a massive theory ($\phi_\ast = g_3/g_4$).  The choice of R-charge also determines the scaling of the operator as the theory flows to the IR.
In investigations  of LG models, one typically assumes such a rescaling has been performed (and the  coefficients of irrelevant deformations have been set to zero) by imposing a quasi-homogeneity condition on the superpotential.

In (0,2) theories there are additional subtleties associated to the $\mu\to 0$ limit of theories with an isolated vacuum.  We will show examples with a seemingly unique choice of scaling of the chiral operators determined by the R-symmetry and the chiral algebra for $\mu \neq 0$, for which the resulting chiral algebra cannot be the chiral algebra of a unitary compact (0,2) SCFT.

The examples we study are (0,2) LG theories.  The UV theory is a family of (0,2) Lagrangians preserving a UV symmetry $\GU(1)^k \times \GU(1)_{\text{R},\tuv}$, with $k\ge 1$ and the first factor describing a global symmetry of the theory.  The interactions are determined by a (0,2) superpotential $\cW$---a polynomial in the chiral superfields that has spin 1/2, is invariant under the global symmetry, and carries UV R-charge $+1$.   A generic theory in the family is obtained by including in $\cW$ all possible monomials in the fields compatible with these symmetries.  We assume that there is a finite number of such monomials, say $P$, and define $\cM_{\tuv} = \C^P$ to be the  ``UV parameter space'' of holomorphic couplings, i.e. the values of the coefficients for the monomials in $\cW$.  There is a discriminant subvariety $\Delta_0 \subset \cM_{\tuv}$ where the scalar potential has non-compact directions in field space.  We are interested in theories with an isolated classical SUSY vacuum, so we restrict the UV parameters to lie in $\cM_{\tuv} \setminus \Delta_0$. 

Fixing a family with $\cM_{\tuv} \setminus \Delta_0$ non-empty, we would like to describe the possible IR fixed points that are obtained via RG flow from theories in this family.  It is clear that $\cM_{\tuv} \setminus \Delta_0$ is a vast over-parameterization of the fixed points.  One source of redundancy is from the action of holomorphic field redefinitions, which induce a $\Delta_0$-preserving action of a group $\cG$ on $\cM_{\tuv}$.  We assume that the D-terms are irrelevant in the sense of RG, and it must then be that every $\cG$-orbit in $\cM_{\tuv} \setminus \Delta_0$ flows to the same IR fixed point.  So, the quotient
\begin{align}
\cM = \{ \cM_{\tuv} \setminus \Delta_0\} / \cG~
\end{align}
should provide a better description of the possible IR fixed points.  There are a number of subtleties associated with this description.  First, $\cM$ will not in general be a separated space, so that the geometric import of the quotient is obscure.  Second, there are in general further IR equivalences:  two distinct orbits can flow to the same fixed point.  This is a feature we encountered in our previous work~\cite{Bertolini:2014ela}.  Finally, there is another possibility:  $\Delta_0$ may not be the true discriminant of the theory, and instead there is a subset $\Delta \subset \cM_{\tuv}$, which may contain $\Delta_0$ as a component, where the IR fixed point fails to be a compact SCFT.  In other words, even if $\cW$ has no flat directions the IR limit may fail to be a compact SCFT unless the UV parameters lie in $\cM_{\tuv} \setminus \Delta$, and it may be that this set is empty for our family of UV theories.

In this work we will consider these structures from the point of view of the chiral algebra of the (0,2) LG theory.  We will assume that a (0,2) LG theory flows to a unitary compact IR SCFT and show that this strongly constrains the form of the chiral algebra if we assume a smooth $\mu\to 0$ limit.  For example, the holomorphic energy-momentum tensor has a unique representative in $H$, and we can use it to find the R-charges of all chiral primary operators.  

We will exhibit examples where the chiral algebra detects the sorts of accidental symmetries studied in~\cite{Bertolini:2014ela}:  more precisely, additional currents, while not symmetries of the Lagrangian, can be found as holomorphic spin--$1$ operators in $H$, and properly including them modifies the Virasoro and Kac-Moody charges of the operators in $H$.  We will also see examples where the chiral algebra cannot have a smooth $\mu \to 0$ limit.  

The rest of the note is organized as follows.  In section~\ref{s:overview} we will set up our conventions for (0,2) Landau-Ginzburg models, review the properties relevant for our analysis, and obtain a number of constraints on the chiral algebra.  Next, in section~\ref{s:examples}, we turn to a number of examples of varying complexity, illustrating the interplay between the UV symmetries, field redefinitions, the chiral algebra, and various unitarity constraints.  In section~\ref{s:terrible} we present  LG theories that do not lie in the naive discriminant $\Delta_0$ but nevertheless cannot flow to a compact unitary SCFT.  We end with a discussion and implications of this result in section~\ref{s:discussion}.   A number of technical computations are relegated to the appendix.

\section*{Acknowledgements}  IVM would like to thank Wesley Engelbrecht and Natasha Gallant for useful conversations; his work is supported in part by NSF grant PHY-1914505.  This work was supported by World Premier International Research Center Initiative (WPI Initiative), MEXT, Japan.

\section{(0,2) Landau-Ginzburg chiral algebra} \label{s:overview}

\subsection{Lagrangian and Symmetries}

\subsubsection*{Notation and conventions}
We work with a flat Euclidean signature worldsheet and (0,2) superspace coordinates $(z; \zb,\theta,\thetab)$, so that the supersymmetry algebra is represented by the superspace derivatives
\begin{align}
\cD & = \frac{\p}{\p\theta} + \thetab \pb~,&
\cDb & = \frac{\p}{\p\thetab} + \theta \pb~,
\end{align}
where $\pb$ is a shorthand $\pb = \frac{\p}{\p\zb}$, and similarly $\p = \frac{\p}{\p z}$.

The UV Lagrangian theories are built with two types of (0,2) chiral superfields:  the bosonic chiral fields $\Phi_a$, with $a=1,\ldots,n$, and the fermionic chiral fields $\Gamma^A$, $A = 1,\ldots,N$, as well as their anti-chiral conjugates $\Phib_a$ and $\Gammab_A$.  Each $\Phi_a$ has lowest component a bosonic field $\phi_a$, while a $\Gamma^A$ has lowest component $\gamma^A$, a Weyl fermion.  By definition, the chiral fields are annihilated by $\cDb$.

The Euclidean action has a standard kinetic term quadratic in the fields, as well as the (0,2) superpotential term $\cW$:\footnote{We lower the index on the anti-chiral fermi multiplets for notational convenience.}
\begin{align}
S & = \frac{1}{2\pi} \int d^2 z \left[ \cD \cDb V_z~ + \ff{1}{\sqrt{2}} \cD \cW + \ff{1}{\sqrt{2}} \cDb\, \cWb \right]~, \nonumber\\
V_z & = -\ff{1}{2} \sum_{a=1}^n  \Phi_a \p \Phib_a -\ff{1}{2} \sum_{A=1}^N \Gammab_A \Gamma^A~, \nonumber\\
\cW & = \sum_{A=1}^{N} \Gamma^A W_A(\Phi)~.
\end{align}

After integrating out the auxiliary fields in the fermi multiplets, we obtain the component action
\begin{align}
S& = \frac{1}{2\pi} \int d^2 z \left[ \sum_{a=1}^n \left( -\phi_a \pb \p \phi_a + \psib_a \p \psi_a\right) + \sum_{A=1}^N \gammab_A \pb \gamma^A 
\right] \nonumber\\
& \quad + \frac{1}{2\pi} \int d^2 z \sum_{A=1}^N \left[ | W_A(\phi) |^2 - \sum_{a=1}^n \left(\gamma^A W_{A,a}(\phi) \psi_a -\gammab_A \overline{ W_{A,a}(\phi)} ~\psib_a \right)\right]~,
\end{align}
where $W_{A,a} = \frac{\p W_A}{\p \phi_a}$~, and the Weyl fermions $\psi_a$ and $\psib_a$ are the superpartners of the bosonic field $\phi_a$.  The action is invariant under (0,2) transformations generated by operators $\bQ$ and $\bQb$ with action on the fields\footnote{Our notation $\bQ \cdot X$ is a shorthand for $\CO{\bQ}{X}$ when $X$ is a bosonic field and for $\AC{\bQ}{X}$ when $X$ is a fermionic field.}
\begin{align}
\label{eq:bigsuperchargetable}
\bQb \cdot \phi_a & = 0~,&
\bQ   \cdot \phi_a & = -\psi_a~,& 
&&
\bQb \cdot \gamma^A & = 0~,&
\bQ  \cdot \gamma^A & = -\Wb_A~,&
\nonumber\\
\bQb \cdot \phib_a & = \psib_a~, &
\bQ   \cdot \phib_a & =  0~, &
&&
\bQb \cdot \gammab_A & = -W_A~,&
\bQ   \cdot \gammab_A & = 0~,& 
\nonumber\\
\bQb \cdot \psi_a & = -\pb \phi_a~, &
\bQ   \cdot \psi_a & = 0~, & 
&&
\nonumber\\
\bQb \cdot \psib_a & = 0~, &
\bQ \cdot \psib_a & = \pb\phib_a~.& 
&&
\end{align}
It is easy to see that $\bQ^2 = 0$, $\bQb^2 = 0$, and, using the $\gamma$ equations of motion, $\AC{\bQ}{\bQb} = \pb$.  

\subsubsection*{Symmetries and charges}
The (0,2) superpotential $\cW$ determines the generators of an ideal $\WW = \la W_1,\ldots, W_N\ra$ in the polynomial ring $\C[\phi_1,\ldots,\phi_n]$.  We restrict attention to quasi-homogeneous ideals, i.e. there is a linear torus action $(\C^\ast)^k$ on the $\phi_a$
\begin{align}
\boldt \cdot \phi_a = (t_1,\ldots,t_k) \cdot \phi_a  =  \phi_a \prod_{\alpha=1}^{k} t_\alpha^{q_a^\alpha}~
\end{align}
specified by the charges $q_a^\alpha$ such that for all $A$ there are charges $Q_A^\alpha$ so that
\begin{align}
W_A(\boldt \cdot \phi)  = W_A(\phi) \prod_{\alpha=1}^{k} t_\alpha^{-Q_A^\alpha}~.
\end{align}
When this is the case the action enjoys a $\GU(1)^k$ global symmetry that assigns charges $q_a^\alpha$ to the $\Phi_a$ and $Q_A^\alpha$ to the $\Gamma^A$ superfields. 
The action also has an R-symmetry $\GU(1)_{\text{R},\tuv}$ with charges $\qb_{\tuv}$ and assignments
\begin{align}
~ && \Phi_a && \Gamma^A && \theta && \bQb ~~\nonumber\\
\qb_{\tuv} && 0 && +1 && +1 && +1~.
\end{align}



Every (0,2) LG theory has a supercurrent multiplet associated to the R-symmetry $\GU(1)_{\text{R},\tuv}$.  In the language of~\cite{Dumitrescu:2011iu} this is the ``$\cR$--multiplet'', with components
\begin{align}
\cR & = \sum_{A=1}^N \Gamma^A\Gammab_A~,\qquad
\cRb   = -\frac{1}{2} \sum_{a=1}^n \cDb\,\Phib^a \cD \Phi^a~,  \nonumber\\
\cT & = - \sum_{a=1}^n \p\Phi^a \p\Phib_a -\frac{1}{2} \sum_{A=1}^N \left(\Gammab^A \p \Gamma^A + \Gamma^A \p \Gammab^A\right)~.
\end{align}
When the ideal $\WW$ is quasi-homogeneous, there are $k$ current multiplets with components
\begin{align}\label{eq:uvcurs}
\cK^\alpha & = \sum_{A=1}^N Q_A^\alpha \Gamma^A \Gammab^A -\frac{1}{2} \sum_{a=1}^n q_a^\alpha\left(\Phi_a \p\Phib_a - \Phib_a \p \Phi_a\right)~, &
\cI^\alpha & =  -\ff{1}{4} \sum_{a=1}^n q_a^\alpha \Phi_a \Phib_a~.
\end{align}
Using the superspace equations of motion
\begin{align}
\cD \Gamma^A & = \sqrt{2}\, \Wb_A~,&
\p \cD \Phi_a & = \sqrt{2} \sum_{A=1}^N \Wb_{A,a} \Gammab^A~
\end{align}
and their conjugates, it is easy to check that
\begin{align}
\pb \cR + \p \cRb & = 0~, &
\cDb \left( \cT - \ff{1}{2} \p \cR\right) & = 0~, \nonumber\\
\pb \cK^\alpha + \p \CO{\cD}{\cDb} \cI^\alpha & = 0~, &
\cDb\left( \cK^\alpha +2 \p \cI^\alpha \right) &  = 0~.
\end{align}
These conditions are the defining relations of the $\cR$-multiplet and current multiplets in (0,2) theories~\cite{Dumitrescu:2011iu,Dedushenko:2015opz,Melnikov:2019tpl}.  The left-hand column is easy to understand: taking the lowest component in the superspace expansion, we just obtain the standard conserved currents of this Lagrangian theory.  

When $k>0$, as we assume, the $\cR$ multiplet is ambiguous:  for any choice of parameters $\tau_\alpha$, we can take
\begin{align}
\label{eq:improvedRmultiplet}
\cR_\tau& = \cR + \sum_{\alpha=1}^k\tau_\alpha \cK^\alpha~,&
\cRb_\tau & = \cRb + \sum_{\alpha=1}^k \tau_\alpha \CO{\cD}{\cDb}\cI^\alpha~,&
\cT_\tau & = \cT - \sum_{\alpha=1}^k \tau_\alpha \p^2 \cI^\alpha~,
\end{align}
and these components will satisfy the defining relations of the $\cR$-multiplet.

We are interested in $\cW$ for which the low-energy dynamics is determined by a compact, unitary (0,2) SCFT with a unique vacuum state.
To this end we restrict the UV parameters to lie in $\cM_{\tuv} \setminus\Delta_0$, so that $\cW$ has the origin as the unique zero of the scalar potential, or, equivalently,  $\WW$ is a zero-dimensional quasi-homogeneous ideal.  Note this is only possible when $N\ge n$.

We will also make a simplifying
technical assumption: linear terms of the form $m_{Ab} \phi_b \subset W_A$ indicate  massive fields that can be integrated out.
Since our interest is in the IR dynamics, we assume this has been done,
and such terms are absent.

\subsection{IR Expectations}

With our assumptions, the symmetries of the IR theory will include symmetries of the UV theory, as well as conformal invariance and
possibly emergent enhanced symmetry.\footnote{Some of the UV symmetries may decouple at low energy, i.e. some UV symmetries may turn out to act trivially in the IR.  We will not encounter this phenomenon in our study of LG theories: every UV symmetry will act non-trivially on the chiral algebra.}   If the SCFT is compact and unitary, the holomorphic and anti-holomorphic components of any conserved current are conserved separately.\footnote{See, for example,~\cite{Ginsparg:1988ui}.}  The right-moving
(anti-holomorphic) currents generate an $\cN{=}2$ superconformal
algebra.  This can be enhanced in several ways:  for example, SUSY can be extended, there can be right-moving symmetry currents, or multiple
 copies of the entire $\cN{=}2$ algebra~\cite{Melnikov:2016dnx}.  We will suppose the SCFT has two supercharges obtained as the low energy limit of the UV supercharges $\bQb,\bQ$, and the corresponding anti-holomorphic supercurrents reside in a superconformal current multiplet with lowest component the right-moving R-current $\Jb_{\text{ir}}$ and top component the right-moving energy momentum tensor $\Tb_{\tir}$.   We will also for the moment assume that $\Jb_{\text{ir}}$ is the only right-moving spin--$1$ current in the SCFT and ignore any possibility of enhancement in the right-moving sector.
 
Unitarity with respect to the right-moving $\cN=2$ algebra  implies well-known
constraints~\cite{Lerche:1989uy} on the charge $\qb$ and the
right-moving Virasoro weight of a state: $\hb\ge |\qb|/2$.  Equality
holds for chiral(antichiral) primary states, annihilated by the supercharge $\bQb$($\bQ$).  The
Sugawara constraint $\hb\ge 3\qb^2/2\cb$ implies that for chiral primary states
$\qb\le\cb/3$.

The left-moving (holomorphic) sector of the SCFT is less constrained.  The
currents include a Kac--Moody--Virasoro (KMV) algebra generated
by the energy-momentum tensor $T_{\tir}(z)$ and a collection of spin--1
holomorphic currents.  Let $\cV_{\tir}$ be the KMV sub-algebra generated by $T_{\tir}$ and the $\GU(1)_{\text{R},\tuv}$--neutral holomorphic currents.  We will also denote by $\cV^{\circ}_{\tir} \subset \cV_{\tir}$ the KMV sub-algebra consisting of $T_{\tir}$ and the currents $J^m$ that span the maximal abelian ideal in $\cV_{\tir}$.

Unitarity of the KMV representation implies that the OPE of the currents in $\cV^\circ_{\tir}$ is 
\begin{align}
J^m(z) J^n(w)\sim \frac{M^{mn}}{(z-w)^2}~,
\end{align}
where $M$ is positive definite, and the charges $q_m$ and
conformal weight $h$ of any state satisfy
$h\ge (q^T M^{-1} q)/2$.  
 
  The LG theory exhibits an
R-symmetry along the flow determined by $\GU(1)_{\text{R},\tuv}$,
associated to the conserved current $J_{\text{R},\tuv}$ given by the
lowest component of the $\cR$-multiplet.  $J_{\text{R},\tuv}$ flows to some
conserved current in the SCFT, and if $\Jb_{\tir}$ is the unique right-moving spin--$1$ current, the difference
$J_{\text{ir}}= \Jb_{\text{ir}} - J_{\text{R},\tuv}$ is a
distinguished left-moving (holomorphic) current.  In particular, the
charges $(q,\qb)$ of all states under
$(J_{\text{ir}},\Jb_{\text{ir}})$ satisfy $q-\qb = \qb_{\tuv} \in\Z$.
We can choose a basis for the currents so that the distinguished
symmetry $J_{\tir}$ has zero two-point functions with the remaining
$k-1$ currents, and 
\begin{align}
\label{eq:Jirnorm}
J_{\tir} (z) J_{\tir}(w) \sim \frac{r}{(z-w)^2}~.
\end{align}
Unitarity implies that the ``level'' $r$ is positive; its value is fixed by the normalizations of $J_{\text{R},\tuv}$ and $\Jb_{\text{ir}}$.

It may seem that the existence of the privileged holomorphic current $J_{\tir}$ depends on our assumption that $\Jb_{\tir}$ is the unique anti-holomorphic spin--$1$ current.  We will show at the level of the chiral algebra that the result is more general:  there is a current $J_{\tir} \subset \cV^\circ_{\tir}$ such that the R-charges of all chiral primary operators are determined by $\qb_{\tir} = q_{\tir} + \qb_{\tuv}$.

\subsubsection*{Spectral flow and holomorphic fields}

As we have seen, 
the charges with respect to $J_{\tir},\Jb_{\tir}$ satisfy  $q_{\tir}-\qb_{\tir} \in\Z$, and therefore the SCFT spectrum includes the
operator $\Sigma$ that generates non-chiral spectral flow by one unit~\cite{Lerche:1989uy,Vafa:1989xc}.
Bosonizing the currents as
\begin{align}
J_{\tir}(z) &= i\sqrt{r}\,\p\varphi(z)~,& 
\Jb_{\tir} &= i\sqrt{\cb/3}\,\pb\varphib(\zb)~,
\end{align}
we can represent $\Sigma$ and its conjugate $\Sigma^\dag$ by\footnote{In appendix~\ref{app:Koszul} we make some comments on representing $\Sigma$ in the chiral algebra of LG theories.}
\begin{align}
\Sigma &= e^{-i\left(\sqrt{r}\,\varphi + \sqrt{\cb/3}\,\varphib\right)}~,&
\Sigma^\dag & = e^{i\left(\sqrt{r}\,\varphi + \sqrt{\cb/3}\,\varphib\right)}~.
\end{align}
Acting on the vacuum by $\Sigma^\dag$ produces a chiral state with $(q_{\tir},\qb_{\tir}) =
(r,\cb/3)$ and weights $(h,\hb) = (r/2,\cb/6)$.  This is the unique state with these charges and weights:  a multiplicity of such states would imply a multiplicity of vacua.

Chiral primary fields with charge $\qb_{\tir} = \cb/3$ are related to
holomorphic fields.  A primary field $\Psi$ with charges $(q_{\tir},\cb/3)$ and weights
$(h,\cb/6)$ can be written as
\begin{align*}
\Psi =  e^{i\left(\frac{q_{\tir}}{\sqrt{r}}\varphi + \sqrt{\cb/3}\varphib\right)}
X\,~,
\end{align*}
where $X$ has trivial OPEs with $J_{\tir}$ and $\Jb_{\tir}$, so that the OPE
\begin{align*}
\Sigma(z,\zb) \Psi(w,\wb) \sim (z-w)^{-q_{\tir}}(\zb-\wb)^{-\cb/3}
\underbrace{e^{i\frac{q_{\tir}-r}{\sqrt{r}}\varphi} X (w)}_{=\Psi'(w)}
\end{align*}
yields a holomorphic field $\Psi'$ of charges $(q_{\tir}-r,0)$ and weights
$(h-q_{\tir}+r/2,0)$. 

\subsubsection*{OPE of chiral primary fields}
More generally, for all chiral primary fields denoted by $\cO_i$, whether holomorphic or not, the OPE takes the form
\begin{align}
\cO_i(z,\zb) \cO_j(0) = \sum_{k} C_{ij}^{~~~k} z^{s_k-s_i-s_j} \cO_k(0) + \bQb(\cdots)~,
\end{align}
where $s_i$ is the spin of $\cO_i$.  As we discuss next, we will be
able to study this structure as an algebra on the cohomology of $\bQb$
and compute it in terms of the UV fields of the LG theory.  This will
be our main tool for translating the constraints imposed by unitarity of
the IR theory to conditions on $\WW$ beyond the two we have already imposed,
namely that $\WW$ should be quasi-homogeneous and zero-dimensional.

\subsection{The chiral algebra of (0,2) Landau-Ginzburg theories}
%
The supersymmetry algebra implies that the cohomology
class $[\cO(z,\zb)] \in H$ of a $\bQb$-closed operator is
$\zb$-independent. This means the OPE of $\bQb$-closed operators
descends to a holomorphic OPE in cohomology.  The position
dependence is determined by Lorentz invariance and can
be expressed in terms of the spins of the operators: 
\begin{align}
[\cO_i(z)] [\cO_j(0)] = \sum_{k} C_{ij}^{~~k}  z^{s_k - s_i-s_j} [\cO_k(w)]~.
\end{align}
This is the chiral algebra\footnote{The {\sl chiral algebra\/} of a CFT was introduced in \cite{Moore:1988qv} to mean the OPE algebra of the holomorphic fields.  In our context it has become customary to use the term to denote a different vertex operator algebra, the OPE algebra of $\bQb$ cohomology.  We use the term in the latter sense.}
  of the theory, and when the end-point of
the RG flow is a compact SCFT, we expect this to coincide with the
algebra of chiral primary operators in the SCFT.  In studying the OPE
we will work exclusively in cohomology in what follows, and where
confusion is unlikely we will conflate a $\bQb$-closed operator with its
cohomology class. 

While the algebra on $H$ can be defined for any QFT (and then
compared to the SCFT that emerges at low energy), it is particularly
computable in LG theories with isolated vacua:  the UV theory is
super-renormalizable, ensuring that the algebra has an exact
presentation in terms of free fields 
$\phi_a$, $\rho_a = \p \phib_a$, $\gamma^A$ and $\gammab_A$ with OPEs
\begin{align}
\label{eq:OPE}
\phi_a(z) \rho_b(w) & \sim \frac{\delta_{ab}}{z-w}~,&
\gamma^A(z)\gammab_B(w) & \sim \frac{\delta^A_B}{z-w}~
\end{align}
and supercharge
\begin{align}
\label{eq:Qbarcoho}
\bQb &= -\oint \frac{dz}{2\pi i} \cJ(z)~, & \cJ(z) & =  \sum_{A=1}^N \gamma^A W_A(\phi)~.
\end{align}
Using this presentation it is conceptually straightforward to
construct representatives for cohomology classes in $H$.
Note that $H$ can be graded by spin $s \in \ff{1}{2}\Z_{\ge 0}$ and the UV symmetries $\GU(1)_{\text{R},\tuv} \times \GU(1)^k$:
\begin{align}
 H = \oplus_s H^s =  \oplus_{s,\qb_{\tuv}} \,H^{s,\qb_{\tuv}} =  \oplus_{s,\qb_{\tuv},\bq}\, H^{s, \qb_{\tuv},\bq}~.
\end{align}
This will prove  useful in our calculations.

\subsection{Holomorphic field redefinitions}
The free field presentation inherited from the Lagrangian theory gives a concrete way to evaluate the chiral algebra, but it is not canonical.  Indeed, for the purposes of computing the chiral algebra it may well be that another choice of fields is more convenient.  In order to commute with the (0,2) supersymmetry action and $\GU(1)_{\text{R},\tuv}$, such a field redefinition should be holomorphic, i.e. take the form

\begin{align}
\phi'^a  & = f^a(\phi_1,\ldots,\phi_n)~, &
\gamma'^A & = \sum_{B=1}^N F^A_B(\phi_1,\ldots,\phi_n) \gamma^B~.
\end{align}
Holomorphy places strong conditions on invertibility:  the transformation is invertible if and only if
\begin{align}
\det_{a,b}  \frac{\p \phi'^a}{\p \phi^b} &= k_1~,&
\det_{A,B} F^A_B & = k_2~
\end{align}
for some non-zero field-independent constants $k_1$ and $k_2$.     When the transformation is invertible, we call it a holomorphic field redefinition.  Setting
\begin{align}
\gammab'_A & = \sum_{B=1}^N (F^{-1})_A^B \gammab_B~, &
\rho'^a & =~ \sum_{c=1}^n : \frac{\p\phi^c}{\p \phi'^a} \left(\rho_c + \sum_{C,D=1}^N A^{D}_{cC}\gamma^C \gammab_D\right) :~,
\end{align}
with
\begin{align}
A^D_{cC} = \sum_{E=1}^N (F^{-1})^D_E \frac{\p}{\p\phi^c} F^E_C~,
\end{align}
we find that $\phi'_a, \rho'^a, \gamma'^A,\gammab'_A$ have the standard OPEs as in~(\ref{eq:OPE}), and setting
\begin{align}
W'_B(\phi') & = \sum_{A=1}^N (F^{-1})^A_B W_A (\phi(\phi'))~,
\end{align}
the supercharge takes the form
\begin{align}
\bQb & =- \oint \frac{dz}{2\pi i} \sum_{A=1}^N \gamma'^A W'_A(\phi')~.
\end{align}
Any two theories related by a holomorphic field redefinition have isomorphic chiral algebras.\footnote{This structure plays an important role in curved $bc$--$\beta\gamma$ systems that play a role in pure spinor strings~\cite{Nekrasov:2005wg}, or in the chiral algebra of (0,2) NLSMs or hybrid theories~\cite{Witten:2005px,Bertolini:2013xga,Bertolini:2017lcz,Bertolini:2018now}.  In those cases there are $bc$--$\beta\gamma$ systems in each patch of the targetspace, and the systems are related by holomorphic redefinitions of this form on overlaps.  The triviality of the targetspace is a significant simplification of the LG setting.}

\subsection{Holomorphic cohomology and symmetries in LG models}
The holomorphic fields of the SCFT are chiral primary with $\qb = 0$, and they form a subalgebra of the algebra of chiral primary operators.   The conjugate (in the sense of producing identity via the OPE) field of any holomorphic field must also be holomorphic, and this leads to strong unitarity constraints on the chiral algebra.  The holomorphic subalgebra includes the KMV algebra  $\cV_\tir$ defined above, which will be our primary focus in the following. 

Some of the symmetries in $\cV_{\tir}$ are inherited from the UV theory: the current multiplets $(\cK^\alpha,\cI^\alpha)$ of~(\ref{eq:uvcurs}) generate symmetries
present along the flow.  The superfields
\begin{align}
\cK_\chi^\alpha = \cK^\alpha + 2 \p \cI^\alpha
\end{align}
are chiral, and their lowest components ( normal ordered with respect to~(\ref{eq:OPE}) )
\begin{align}
\label{eq:UVcurrents}
J^\alpha = \sum_{A=1}^N Q_A^\alpha :\gamma^A\gammab_A: - \sum_{a=1}^n q_a^\alpha :\phi_a \rho_a :
\end{align}
determine $\bQb$-closed, $\GU(1)_{\text{R},\tuv}$ neutral spin--$1$ operators.  The OPE of these is
\begin{align}
\label{eq:UValgebra1}
J^\alpha(z) J^\beta(w) \sim M^{\alpha\beta}(z-w)^{-2}~,
\end{align}
where the $k\times k$ symmetric matrix $M^{\alpha\beta}$ has components
\begin{align}
M^{\alpha\beta} & =\sum_{A=1}^N Q_A^\alpha Q_A^\beta - \sum_{a=1}^n q_a^\alpha q_a^\beta~.
\end{align}
$M$ measures the mixed anomaly between the two currents.
We assume, as is the case in all of our examples, that this is
positive definite.  Because $M$ is also field-independent and therefore R-neutral, each $J^\alpha$ must be R-neutral as well, and that means all of the currents $J^\alpha$ descend
to nontrivial commuting holomorphic currents that belong to $\cV_{\tir}$.  We denote the KMV sub-algebra generated by $T_{\tir}$ and the currents $J^\alpha$ by $\cV_{\tuv}$.

The IR theory may enjoy emergent symmetries in addition to
those already present in the UV.  In a compact unitary SCFT each of the continuous symmetries should be anti-holomorphic or holomorphic, and each of the holomorphic currents should have a representative in the chiral algebra.
We will now describe all such currents that are neutral with respect to $\GU(1)_{\text{R},\tuv}$.  

\subsubsection*{Holomorphic currents}
A spin--$1$ operator in $H$ neutral with respect to $\GU(1)_{\text{R},\tuv}$ must take the form 
\begin{align}
J =  \sum_{a=1}^n :A^a(\phi)\rho_a: +  \sum_{A,B =1}^N B_A^B(\phi) :\gamma^A\gammab_B: + \sum_{a=1}^n C^{a}(\phi) \p\phi_a~,
\end{align}
where the coefficients $A$, $B$, and $C$ are further constrained by $\bQb \cdot J = 0$.
We will now show that the last term must be absent, i.e. $C^a = 0$ for all $a$ if the chiral algebra describes the chiral primary states of a unitary compact SCFT.  

The current $J$ must be neutral under the IR R-symmetry generated by $\Jb_{\tir}$.  This R-symmetry must act linearly on the space of spin--$0$ chiral primary fields, and in the LG chiral algebra this space is isomorphic to the quotient ring
\begin{align}
R_{\WW} = \C[\phi_1,\ldots,\phi_n] / \WW~.
\end{align}
Unitarity implies that the R-symmetry action on $R_{\WW}$ must be diagonalizable, and the action must respect the OPE structure, which is simply multiplication in the ring $R_{\WW}$.  As we show in Appendix~\ref{app:Rsym}, our assumptions on the R-symmetry action and the ideal imply that we can find a holomorphic field redefinition to obtain $n$ generators $\phi'_a = f_a(\phi_1,\ldots,\phi_n)$ of $R_{\WW}$ that transform linearly under the R-symmetry:
\begin{align}
\delta \phi'_a = i \alpha \qb_a \phi'_a~,
\end{align}
where $\alpha$ is the infinitesimal parameter.  Unitarity further implies that $\qb_a \ge 0$, and compactness excludes $\qb_a = 0$:  any spin--$0$ chiral field with $\qb_a = 0$ would be a non-identity operator of zero dimension and spin.   It follows that any R-neutral current $J$ cannot have a term of the form $C_a(\phi') \p \phi'_a$, and in terms of the original fields this implies that $C^a(\phi)=0$ in $J$.

So, the most general
form of a holomorphic $\GU(1)_{\text{R},\tuv}$-neutral current is of the form
\begin{align}\label{eq:genspin1}
  J[A,B] = \quad \sum_{a=1}^n:A^a(\phi)\rho_a: + \sum_{A,B=1}^N B_A^B(\phi) :\gamma^A\gammab_B:~.
\end{align}
%
Using~(\ref{eq:Qbarcoho}) we find
\begin{align}
\label{eq:QbardotJ}
  \bQb\cdot J = -\sum_{A=1}^N\left( \sum_{a=1}^n A^aW_{A,a} - \sum_{B=1}^N B_A^B W_B\right)\gamma^A~.
\end{align}
Thus $J$ is closed, i.e.
\begin{align}
\label{eq:Jclosed}
 -\sum_{a=1}^n A^aW_{A,a} + \sum_{B=1}^N B_A^B W_B = 0~,
\end{align}
if and only if $\cW$ is invariant under the infinitesimal transformation
\begin{align} \label{eq:Wsymmetry}
  \Phi_a\to\Phi_a - \epsilon A^a(\Phi) \quad \Gamma^A\to\Gamma^A +
  \epsilon \sum_{B=1}^N B_B^A(\Phi) \Gamma^B~.
\end{align}
The solutions to these linear equations in the coefficients of the
polynomials $A_a$ and $B_A^B$ determine the space of all
$\bQb$-closed spin--$1$ fields.  We denote by $J^M$ a basis for this space. 

Not all of these correspond to conserved holomorphic currents.  There is a subspace associated to $\bQb$-trivial solutions: these are symmetries that act trivially in  the IR theory.  In addition, it is
possible that some of solutions to~(\ref{eq:Jclosed}) determine spin--$1$
fields of {\sl positive\/} R-charge despite  being neutral under the
UV symmetry.   An expedient way to detect which of the solutions
determine true IR symmetries is to compute the OPE
\begin{align}
  J^M(z)J^N(w)\sim \frac{\Mh^{MN}}{(z-w)^2} + O(z-w)^{-1}~,
\end{align}
where the symmetric matrix $\Mh$ has components
\begin{align}
\Mh^{MN} & = \sum_{A,B=1}^N (B^M)^B_A (B^N)_B^A - \sum_{a,b=1}^n \p_a (A^M)^b \p_b (A^N)^a~.
\end{align}
The coefficients $\Mh^{MN}$ will in general be field-dependent, but from our discussion of the R-symmetry action, it is clear that such a term cannot arise from the OPE of two R-neutral currents.  So, we see that the kernel of $\Mh|_{\phi = 0}$ will contain exact currents (which have nonsingular OPE with all closed fields) as well
as R-charged fields (which cannot produce the identity under OPE with
any chiral field).  Projecting out the kernel of $\Mh|_{\phi=0}$ is thus an
expedient way to determine the full holomorphic symmetry of the IR theory that commutes with $\GU(1)_{\text{R},\tuv}$: these are the currents in $\cV^{\tir}$.

To obtain the abelian currents in $\cV^{\circ}_{\tir}$, we just further restrict this sector to the abelian holomorphic currents labeled by $J^m$, with OPE given by the positive-definite matrix $M^{mn}$.  In practice, to obtain $\cV_{\tir}^\circ$ we can restrict to currents $J^M$ that commute with the currents $J^\alpha$ of $\cV_{\tuv}$, and  this leads to a significant simplification in finding this smaller symmetry algebra.

\subsubsection*{General holomorphic symmetry currents for $N=n$}
A holomorphic current $J$ may in principle have terms charged with respect to $\GU(1)_{\text{R},\tuv}$.  However, we show in appendix~\ref{app:GenSym} that when $N=n$ every holomorphic current still takes the form given in~(\ref{eq:genspin1},\ref{eq:Jclosed}) and is in fact $\GU(1)_{\text{R},\tuv}$--neutral.

\subsection{The KMV algebra in cohomology}

The fact that there can be accidental symmetries in the IR theory, and
that these can, by mixing with the UV R-symmetry, invalidate
predictions about the IR theory based on the UV symmetries has been
known for a long
time~\cite{Benini:2012cz,Melnikov:2016dnx,Melnikov:2019tpl}.  What we
add to the discussion is, in the restricted context of (0,2) LG
models, an explicit calculation of the full IR symmetry.   To
determine the IR R-symmetry, we determine the left-moving
energy-momentum tensor $T_\tir$, or more precisely its representative in $H^2$.  The R-charge of any chiral operator
is then determined by its dimension  $h$ with respect to $T_{\tir}$
and its spin $s$:
\begin{align}
\label{eq:qbarfromspin}
\qb_{\tir} =2\hb = 2 (h-s)~.
\end{align}

To determine $T_{\tir}$ we observe that the Virasoro generator $L_0 = \oint \frac{dz}{2\pi i}  z T_{\tir}(z)$ 
must have diagonalizable actions on $H^0$ and $H^{1/2}$, and we will assume that the 
the spin--$0$ fields $\phi_a$ and spin--$1/2$ fields $\gamma^A$ are quasi-primary, so that
\begin{align}
\label{eq:quasiprimary}
T_{\tir} (z) \phi_a(w) &\sim \cdots + \frac{\ff{1}{2} \cA^a(\phi)}{(z-w)^2} + \frac{ \p \phi_a(w)}{z-w}~, \nonumber\\
T_{\tir} (z) \gamma^A(w) & \sim \cdots + \frac{\ff{1}{2} \sum_{B=1}^N \cB^A_B(\phi)\gamma^B(w)}{(z-w)^2} + \frac{\p\gamma^A(w)}{z-w}~,
\end{align}
where the coefficients $\cA^a(\phi)$ and $\cB^A_B(\phi)$, which determine the action of the Virasoro generator $L_0$, are constrained by the symmetries of the theory, including the R-symmetry.

Before we pursue the consequences of~(\ref{eq:quasiprimary}), we point out  that in general we might expect a weaker form of~(\ref{eq:quasiprimary}), where the first-order poles are modified by $\bQb$-exact terms.  When compared to the free field representative of $T_{\tir}$ consistent with~(\ref{eq:quasiprimary}), this leads to additional operators that are expressed in terms of the coefficients of the $\bQb$-exact modifications of the poles, and the coefficients are subject to a set of complicated algebraic constraints involving the ideal generators $W_A$ and their derivatives.  It can be shown that every $\bQb$-exact term in the simple pole of the OPE of $T_{\tir}$ with $\phi^a$ can be removed by shifting $T_{\tir}$ by a $\bQb$-exact operator.   We have not been able to show that the same is true of a $\bQb$-exact term in the simple pole of the OPE with $\gamma^A$, but we suspect that the $\bQb$-exact modifications of the poles may only be possible for a special class of ideals $\WW$.  In contrast, using the stronger form of~(\ref{eq:quasiprimary}) leads to a universal form for $T_{\tir}$ just determined by the holomorphic symmetries of $\cW$.    It may be interesting to pursue the general case further, but, as we will discuss in section~\ref{ss:badunicorn}, any such modification cannot resolve the unitarity puzzles we will encounter. 

Returning now to consider~(\ref{eq:quasiprimary}), we observe that any LG theory has a spin $2$  $\bQb$-closed field
\begin{align}\label{eq:Tzero}
  T_0 =  - \sum_{a=1}^n:\rho_a\p\phi^a: - \sum_{A=1}^N:\gamma^A\p\gammab_A:~.
\end{align}
This is manifestly neutral under the UV symmetries.
Computing the OPE of this with the chiral LG fields, we find
\begin{align}
  T_0(z)\phi^a(w)&\sim \frac{\p\phi^a(w)}{z-w}~,&
  T_0(z)\gamma^A(w)&\sim \frac{\gamma^A(w)}{(z-w)^2} +
  \frac{\p\gamma^A(w)}{z-w}~.
\end{align}
This implies that $T_\tir$ is given by 
\begin{align}
T_\tir = T_0 + T'~,
\end{align} 
where
$T'$ is $\bQb$-closed, neutral under the UV symmetries, and 
\begin{align}
  T'(z)\phi^a(w)&\sim \cdots + \frac{\ff{1}{2} \cA^a(\phi)}{(z-w)^2}~, &
  T'(z)\gamma^A(w)&\sim \cdots + \frac{ \ff{1}{2} \sum_{B=1}^N( \cB^A_B(\phi)-\delta^A_B)\gamma^B(w)}{(z-w)^2}~.
\end{align}
We will now show that~(\ref{eq:quasiprimary}) implies
\begin{align}
\label{eq:Tprimereq}
  T' = \ff{1}{2} \p K~,
\end{align}
where $K = J[\cA,\cB]$ is a $\bQb$-closed, spin--$1$ and $\GU(1)_{\text{R},\tuv}$--neutral current, of the form
described in~(\ref{eq:genspin1},\ref{eq:Jclosed}):
\begin{align}
\label{eq:specialK}
  K(z) = \sum_{a=1}^n :\cA^a(\phi)\rho_a: + \sum_{A,B=1}^N :\cB_A^B (\phi)\gamma^A\gammab_A:~.
\end{align}
We will also prove that $K \in \cV^\circ_{\tir}$ and determine its precise form in terms of the $J^m$.

\subsubsection*{The general form of $T'$}
We begin by writing the most general
possible spin--$2$ operator that is neutral under the R-symmetry and $\GU(1)_{\text{R},\tuv}$:
\begin{align}
  X &=~ :O_1{}^{ab}\rho_a\rho_b: + :O_2{}^a\p\rho_a: +
      :O_3{}^b_a\p\phi^a\rho_b:+ :O_4{}_A^{aB}\rho_a\gamma^A\gammab_B: \nonumber\\
  &\qquad + :O_5{}_{aA}^B\p\phi^a\gamma^A\gammab_B: + :O_6{}_A^B\gamma^A\p\gammab_B: +  :O_7{}_A^B\p\gamma^A\gammab_B:   \nonumber\\
  &\qquad  + :O_8{}_{AB}^{CD} \gamma^A\gamma^B\gammab_C\gammab_D:~,
\end{align}
where $O_i$ are polynomials in $\phi$ of appropriate  degrees.\footnote{To lighten notation we use the summation convention in this section: there is an implied sum on all repeated indices.}  Note that a term of the form $O^{ab} \p\phi_a \p\phi_b$ is excluded by R-symmetry neutrality:  we saw a similar term excluded in our analysis of holomorphic currents given above.

Computing the OPE with $\phi^a$ we find
\begin{align}
  X(z)\phi^a(w)&\sim \frac{O_2{}^a}{(z-w)^2} \nonumber\\
  &\qquad + \frac{\left(\p_b
  O_2{}^a-O_3{}_b^a\right)\p\phi^b - 2 O_1{}^{ab}\rho_b -
  O_4{}_A^{aB}\gamma^A\gammab_B}{z-w}~.
\end{align}
The absence of a simple pole implies $O_1=0=O_4$ and $O_3{}_b^a=\p_b
O_2{}^a$.

Repeating the calculation with $\gamma^A$ we now have
\begin{align}
  X(z)\gamma^A(w)&\sim \frac{O_6{}_B^A\gamma^B}{(z-w)^2}
  + \frac{1}{z-w}\Big[\left(O_5{}_{aB}^A -
    \p_aO_6{}_B^A\right)\p\phi^a\gamma^B
    + \left(O_7{}^A_B-O_6{}^A_B\right)\p\gamma^B  \Big] \nonumber\\
    &\qquad  + \frac{1}{z-w} 2O_8{}^{AB}_{CD}:\gamma^C\gamma^D\gammab_B:~.
\end{align}
The absence of a simple pole implies
$O_{8}=0$, $O_7=O_6$, and $O_5{}_{aB}^A = \p_a O_6{}^A_B$.
Inserting all of these into $X$ and relabeling $O_2{}^a=\cA^a$ and $O_6{}^B_A = \cB^B_A$, we obtain~(\ref{eq:Tprimereq}) and~(\ref{eq:specialK}).
$T_{\tir}$ is $\bQb$-closed if and only if $K$ is, which means the coefficients $\cA^a$ and $\cB^B_A$ generate a symmetry of $\cW$.
%
%
%

\subsubsection*{$K$ belongs to $\cV_{\tir}^\circ$}

A short computation shows that for any $\GU(1)_{\text{R},\tuv}$--neutral current $J[A,B]$
\begin{align}
T_0 (z) J(w) \sim \frac{\sum_{A=1}^N B_A^A - \sum_{a=1}^n\p_a A^a}{(z-w)^3}  + \frac{ J}{(z-w)^2} + \frac{\p J}{(z-w)}~,
\end{align}
and in particular
\begin{align}
T_0 (z) K(w) \sim \frac{\sum_{A=1}^N \cB_A^A - \sum_{a=1}^n\p_a \cA^a}{(z-w)^3}  + \frac{ K}{(z-w)^2} + \frac{\p K}{(z-w)}~.
\end{align}
Since $L_0$ must commute with the R-symmetry transformations of $\phi_a$ and $\gamma^A$, the coefficient of the cubic pole must be constant,\footnote{Equation~(\ref{eq:quasiprimary}) and non-negativity of the R-charges together imply that the individual terms $\p_a \cA^a$ and $\cB^A_A$ must be field-independent:  for example $\cA^a(\phi)$ can at most depend linearly on $\phi_a$. }
 and that in turn shows that $T_0$ is neutral with respect to $K(w)$.  By the same logic, $T_0$ is also neutral with respect to any current $J$ in $\cV_{\tir}$, and R--invariance of $J$ implies that the coefficient of the cubic pole in the $T_0 J$ OPE is field-independent.

 Therefore, since $T_{\tir} = T_0 + \ff{1}{2} \p K$ must be neutral with respect to all holomorphic symmetries, $K$ must commute with all $\GU(1)_{\text{R},\tuv}$--neutral currents and therefore belong to $\cV^\circ_{\tir}$.  Ultimately this is the reason for our focus on this subset of currents:  they are the only ones that participate in determining the energy-momentum tensor.

So, we can expand $K$ in our basis of spin--$1$ currents in $\cV^\circ_{\tir}$ as
\begin{align}
K = \sum_{m} \tau_m J^m
\end{align}
for some constants $\tau_m$. The $\tau_m$ are fixed by computing the OPE of
$T_\tau = T_0 + \ff12\sum_n\tau_n\p J^n$ with the currents $J^m$:
\begin{align}
  T_\tau(z) J^m(w)\sim \frac{\sum_{A=1}^N B^m{}_A^A - \sum_{a=1}^n\p_a A^m{}^a - \sum_p
  \tau_pM^{pm}}{(z-w)^3} + \frac{J^m(w)}{(z-w)^2} + \frac{\p
  J^m(w)}{z-w}~.
\end{align}
This is consistent with the KMV algebra provided $\tau_m$ are given by
\begin{align}\label{eq:pickr}
  \tau_m =  \sum_{n} \left(M^{-1}\right)_{mn}\left( 
  \sum_{A=1}^NB^n{}_A^A-\sum_{a=1}^n\p_a A^n{}^a\right)~.
\end{align}
Setting $\tau$ to these values in $T_{\tau}$, we obtain $T_{\tir}$.

The symmetry $K$ is related to the $J_{\tir}$ symmetry introduced above:  $J_{\tir} = -K$.  To see this, we note that by extending the arguments of appendix~\ref{app:Rsym}, there exists a field redefinition to fields $\phi_a$ and $\gamma^A$ that are simultaneous eigenstates of the R-symmetry and $K$.  Letting $q_a$ and $Q_A$ denote the charges of $\phi_a$ and $\gamma^A$ with respect to $-K$, we find the weights with respect to $T_{\tir}$ to be
\begin{align}
h_a &= \ff{1}{2} q_a~, & h_A & = 1+\ff{1}{2} Q_A~,
\end{align}
and the R-charges are then given by~(\ref{eq:qbarfromspin}):
\begin{align}
\qb_a & = q_a~,&  \qb_A = 1+Q_A~,
\end{align}
and this is consistent with our definition $\Jb_{\tir} = J_{\tir} + J_{\text{R},\tuv}$.

\subsubsection*{$\cV_{\tir}^\circ$ and unitarity constraints}

Thus, we have an algorithmic procedure to find the KMV subalgebra  $\cV_{\tir}$ of the chiral algebra:  (i) compute the space of $\bQb$-closed, spin--$1$, $\GU(1)_{\text{R,\tuv}}$-neutral operators, project out the kernel of the matrix 
$\Mh|_{\phi=0}$; (ii) compute the improved energy momentum tensor $T_{\tir}$ by evaluating the coefficients $\tau_m$.  This structure already encodes a great deal of information about the IR SCFT.

More generally, we can in principle study the full chiral ring of the SCFT.   This algebraic structure will in general depend on any marginal parameters that can be encoded as coefficients in the ideal generators $W_A$, and the detailed structure goes beyond coarser invariants such as the elliptic genus of the theory.  The latter is still a useful characteristic of the theory, and can be computed explicitly in gauged linear sigma models and LG theories~\cite{Gadde:2013dda,Benini:2013xpa,Benini:2013nda,Gadde:2016khg}.

What is perhaps equally remarkable is that the chiral algebra is completely determined by a choice of $\WW$, and so far we placed rather weak restrictions on it.  Yet we also know that the algebra of chiral primary operators of a compact SCFT is constrained by various unitarity requirements.  Some of the most basic ones are:
\begin{enumerate}
\item  $\cb \ge 0$;
\item  for any chiral primary operator the R-charge must satisfy $ 0 \le \qb \le \cb/3$;
\item the leading term in the OPE of holomorphic spin--$1$ currents must be positive.
\end{enumerate}
It is by no means obvious that these conditions hold for all $\WW$ that satisfy the assumptions spelled out in the previous section.  The main point of our work is to illustrate that there are cases when these conditions are not satisfied.

%

\section{Using the chiral algebra to diagnose the IR} \label{s:examples}

In this section we provide some examples of how the chiral algebra can
be used to study properties of the IR theory.   These demonstrate the
power of the method, and the way it overcomes the limitations of
attempts to use the UV properties directly to infer the IR
dynamics. In the next section, however, will show that this too has
limitations, because in general the chiral algebra of a LG theory in our class cannot be isomorphic to the chiral sector of a compact unitary SCFT.

\subsubsection*{The UV starting point}

Our starting point is a Lagrangian formulation for the LG theory specified by a zero-dimensional ideal $\WW$ with $k\ge 1$ quasi-homogeneity relations.  If we assume that the IR theory does not exhibit an accidental symmetry, then we can construct a candidate KMV algebra $\cV_{\tuv}$ by taking $J_{\tir} = \sum_{\alpha=1}^{k} \tau_\alpha J^\alpha$, where the currents $J^\alpha$ are given in~(\ref{eq:UVcurrents}).  

A remarkable feature of the chiral algebra structure is that it gives us an in-principle method to find conditions on the parameters of a family of ideals $\WW$ that imply an accidental symmetry, e.g. the proper containment $\cV_{\tuv} \subset \cV_{\tir}$.  In general it is not so easy to describe these conditions explicitly, so it is useful to have some simpler diagnostics that do not involve the full computation of $\cV_{\tir}$.  One such diagnostic is obtained by applying the unitarity constraints to representations of $\cV_{\tuv}$.  When these are violated, one resolution is that $\cV_{\text{ir}} \not\simeq \cV_{\text{uv}}$, and we can test for this resolution simply by computing $\cV_{\text{ir}}$.

The reader may ask why she should bother with $\cV_{\text{uv}}$ at
all:  after all, given an ideal $\WW$, the ``true'' KMV algebra
$\cV_{\text{ir}}$ can be obtained explicitly.  There are two
reasons:  first, while describing $\cV_{\text{ir}}$ is conceptually
straightforward, it is not so easy to construct it explicitly, even in
examples with a small number of fields and monomials in the $W_A$;
second, by studying cases with $\cV_{\text{uv}} \not\simeq
\cV_{\text{ir}}$, we access the low energy physics of a strongly
coupled field theory with accidental symmetry:  this may be an
academic topic, but it is surely one of intrinsic interest.  

Our approach to the various examples we consider will be from this point of view: we will first study $\cV_{\tuv}$ and compare it to $\cV_{\tir}$ as necessary.

\subsection{The $k=1$ case}
%
%
%

We begin with the generic LG theory compatible with our assumptions, where there is just a single quasi-homogeneity relation constraining the ideal $\WW$.  In this case the single parameter $\tau$ can be absorbed into a normalization of the charges, and~(\ref{eq:pickr}) reduces to fixing this normalization so that
\begin{align}
-\sum_{A=1}^N Q_A (1+Q_A) = \sum_{a=1}^n q_a (1-q_a)~.
\end{align}
Once this holds, the central charges are given by
\begin{align}
\label{eq:centralcharges}
\frac{\cb}{3} &= \sum_{a=1}^n (1-q_a) - \sum_{A=1}^N (1+Q_A)~,&
c & = N-n+\cb~,
\end{align}
and
\begin{align}
\label{eq:level}
r = \sum_{A=1}^N Q_A^2 - \sum_{a=1}^n q_a^2 = -\sum_{A=1}^N Q_A -\sum_{a=1}^n q_a
\end{align}
is the normalization of $J_{\tir}$ as in~(\ref{eq:Jirnorm}).

Now using $T_{\tir}$ and $J_{\tir}$ OPEs with the fields $\phi_a$, $\rho_a$, $\gamma^A$, and $\gammab_A$, we obtain the following charge and weight assignments:
\begin{align}
\label{eq:charges}
\text{field} && \phi_a && \rho_a 		&& \gamma^A  && \gammab_A \nonumber\\
\text{spin} && 0		&& 1			&&1/2		&&1/2\nonumber\\
q_{\tir}		&&  q_a	&& -q_a		&& Q_A	        && -Q_A	 \nonumber\\
h		&& q_a/2	&&1-q_a/2	&& 1 +Q_A/2    && -Q_A/2      \nonumber\\
\qb_{\tir}	&& q_a  && -q_a		&& 1+Q_A	&&-1-Q_A 
\end{align}
The last row was obtained by using~(\ref{eq:qbarfromspin}).  Not all of these fields belong to $\bQb$ cohomology, but we can use these assignments to determine the charges and weights of operators in $H$ that are constructed as normal-ordered products of these fields.


With some small modifications, this analysis also applies to the $k>1$ setting~\cite{Benini:2012cz,Melnikov:2016dnx}:  there is a preferred linear combination $J_{\tir} = \sum_\alpha \tau_\alpha J^\alpha$ that combines with the naive R-symmetry of the Lagrangian to yield a candidate R-symmetry in the IR, and the preceding discussion applies to that symmetry.  The remaining $k-1$ abelian symmetries can be taken to be orthogonal to $J_{\tir}$.  We can also grade $H$ with respect to these, and there will be additional unitarity conditions on the spectrum imposed by the presence of these symmetries.  Finally, there may also be non-abelian symmetries of the Lagrangian theory, and they can in principle be handled by analogous techniques.

\subsubsection*{Unitarity constraints for $k=1$}
If $\cV_{\text{uv}} \simeq \cV_{\text{ir}}$, it must be that $\cb$ and $r$ calculated from~(\ref{eq:centralcharges},\ref{eq:level}) are positive.  Using~(\ref{eq:charges}), we see that there are further constraints.  For example, the fields $\phi_a$ and $\gamma^A$ are always non-trivial in $H$.\footnote{This is ensured by the absence of linear terms in the $W_A$.}   Therefore, it must be that
\begin{align}
\label{eq:basicbounds}
0 \le q_a \le \cb/3~,&&
0 \le 1+Q_A  \le \cb/3~.
\end{align}
We can obtain stronger bounds on the $q_a$ by invoking compactness.  First, we need $q_a >0$, since otherwise we have a spin--$0$ field with zero R-charge---a signature of non-compactness.  Furthermore,
in order for the ideal $\WW$ to be zero-dimensional, a necessary condition is that every field $\phi_a$ appear with a ``pure'' monomial $\phi_a^{k_a}$ in some generator $W_A$.\footnote{If this is not the case, there is a non-compact locus where $W_A = 0$ for all $A$ obtained by setting all $\phi_b = 0$ with $b\neq a$.  This is familiar from the study of local rings in commutative algebra; see, e.g.~\cite{Cox:1998ua}.}  Since we are not interested in theories with mass terms, it must be that $k_a\ge 2$, and $1+Q_A = 1 - k_a q_a$.  So, the second inequality in~(\ref{eq:basicbounds}) now implies $q_a \le 1/2$.  

These are just special cases of the more general requirement mentioned above: unitarity and~(\ref{eq:qbarfromspin}) imply that  for any $\cO \in H$ with spin $s_\cO$ and weight $h_\cO$, it must be that
\begin{align}
2(h_\cO - s_\cO) \le \cb / 3~.
\end{align}
Given an ideal $\WW$ with one quasi-homogeneity relation, it is in principle possible to compute $\sup_{\{\cO\}} ( h_\cO -s_\cO)$, which would give the most stringent unitarity bound that we can obtain in this fashion.  This does not appear to be easy and is most likely model-dependent.  However, there is a universal operator that appears in $H$ for every LG theory satisfying our assumptions.  This is the operator
\begin{align}\label{eq:unicorndef}
\cU & = \gamma^1 \gamma^2 \cdots \gamma^N \in H^{N/2}~,
\end{align}
with spin $s_{\cU} = N/2$ and $\qb_{\cU} = \sum_{A=1}^N (1+Q_A)$.  $\cU$ is obviously $\bQb$-closed.  To be $\bQb$-exact, it must arise as $\bQb\cdot Y$, where $Y$ is constrained by $\GU(1)_{\text{R},\tuv}$ and spin to be of the form
\begin{align}
Y = \sum_{A_1,\ldots,A_{n-1}=1}^N : f^b_{A_1\cdots A_{n-1}}(\phi) \gamma^{A_1}\cdots\gamma^{A_{n-1}} \rho_b :
+ \sum_{A=1}^N  :g_A(\phi) \gamma^1\cdots \gamma^N \gammab_A :~.
\end{align}
But, since
\begin{align}
\bQb \cdot \rho_b &=- \sum_{A=1}^N \gamma^A W_{A,b}~, &
\bQb \cdot \gammab_A & = - W_A~,
\end{align}
and each $W_{A,b}$ is $\phi$-dependent (recall that we exclude mass terms), $\cU$ cannot be in the image of $\bQb$.\footnote{We also checked in numerous examples that the presence of $\cU \in H^{N/2}$ is consistent with the elliptic genus of the theory.} In what follows, we will refer to $\cU$ as the unicorn operator.

Since it is always present, if the SCFT is compact and $\cV_{\text{uv}} \simeq \cV_{\text{ir}}$, then
\begin{align}
\label{eq:unicornbound}
\sum_{A=1}^N (1+Q_A) \le \cb/3~.
\end{align}
This is a strong requirement.  For example,~(\ref{eq:unicornbound}), (\ref{eq:basicbounds}),~(\ref{eq:centralcharges}), and $q_a \le 1/2$ lead to a bound on the number of bosonic fields in terms of the central charge:
\begin{align}
n \le \begin{cases}  \cb & \text{if}~~ N=n~, \\  4 \cb/3 & \text{if} ~~N>n~. \end{cases}
\end{align}
It is known~\cite{Kreuzer:1992bi,Klemm:1992bx} that in LG theories with (2,2) supersymmetry, where $N=n$ and $W_{a,b} = W_{b,a}$, the bound $n\le \cb$ holds for any zero-dimensional ideal; this bound plays an important role in classifying (2,2) LG theories at fixed central charge.  In (0,2) theories it is possible to choose quasi-homogeneous zero-dimensional ideals consistent with~(\ref{eq:basicbounds}) but violating~(\ref{eq:unicornbound}): there exist infinite families of such ideals with a fixed value of $\cb$. For example, we can set $N=n=3$ and take the following family labeled by a non-negative integer $p$:
\begin{align}
\label{eq:infinitefamily}
W_1 &  = \phi_1^5 +\cdots~,&
W_2 & = \phi_2^{2+p} + \phi_1^6 \phi_2^{1+p} + \cdots~,&
W_3 & = \phi_3^2 + \cdots~,
\end{align}
with $q_2 = 6 q_1$ and $q_3 = (8+6p) q_1$.  A generic choice of coefficients leads to $k=1$ and $\cb = 1$ for any integer $p$.  On the other hand,
\begin{align}
\sum_{A=1}^3 (1+Q_A) = 3-\frac{11 + 6p}{6+4p} \ge \frac{3}{2} > \frac{\cb}{3}~.
\end{align}
The existence of such families is one reason why there is not a straightforward generalization of the (2,2) LG classification results to the (0,2) setting.

The theories of~(\ref{eq:infinitefamily}) have a seeming pathology since $\cU$ violates the unitarity bound, but this is an ultraviolet delusion: there are accidental symmetries for all values of parameters, and $J_{\tir} \not\in \cV_{\tuv}$.  By finding $J_{\tir} \in \cV_{\tir}^\circ$ and computing the correct charges, we find that all unitarity bounds are satisfied and the IR fixed point is a product of three (2,2) minimal models. In what follows we will study a number of such examples.

\subsection{Accidental symmetries and chiral algebra: simple examples}

\subsubsection*{The unicorn operator in a minimal model}
Consider a LG theory that flows to an $A_{k}$ (2,2) minimal model with central charge $\cb/3 = k/(k+2)$.  This has a (0,2) Lagrangian with $N=n$ and $W= \phi^{k+1}$.  The unicorn operator is  $\cU = \gamma$, and its R-charge $\qb = 1/(k+2)$ satisfies the bound $\qb \le \cb/3$.   More generally, we consider operators of the form 
\begin{align}
\cU_l = \gamma \p \gamma \p^2\gamma \p^3 \gamma \cdots \p^{l-1} \gamma~~.
\end{align}
These remain $\bQb$-closed, but such an operator with more than $k$ $\gamma$ insertions will violate unitarity.   How does the minimal model avoid this?  The explanation is simple:  all of the putative non-unitary operators are trivial in cohomology.  Observe that 
\begin{align}
\bQb \cdot \rho &= -(k+1)\gamma\phi^k~,&
\bQb \cdot \gamma \rho^2 & = -k(k+1) \phi^{k-1} \gamma \p \gamma~.
\end{align}
Continuing in the same way, we find that all operators of the form
\begin{align}
\cU_m = \phi^{k+1-m} \gamma \p\gamma \p^2\gamma \cdots \p^{m-1} \gamma
\end{align}
are $\bQb$-exact, and setting $m=k+1$, we see that all $\cU_l$ with $l\ge k$ are $\bQb$-exact.

The example illustrates a mechanism by which some potential violations of unitarity are avoided in the chiral algebra: the operators that would naively lead to violation of unitarity are not in cohomology.  This is a familiar feature from studies of the (c,c) ring of (2,2) LG theories, as well as the B/2 ring of (0,2) LG theories.  We make some comments on these features in appendix~\ref{app:Koszul}.

\subsection*{Accidental symmetry for an irrelevant deformation}
We now give an example where an accidental symmetry explains the seeming non-unitarity of the unicorn operator.  Our test-case has $N=n =2$ with ideal
\begin{align}
\WW = \la \phi_1^2~, \phi_2^2 + \phi_2 \phi_1^2\ra~.
\end{align}
This ideal has $k=1$, and if we assume $\cV_{\tuv} = \cV_{\tir}$, we obtain
\begin{align}
J_{\tir}^{\text{naive}} = -\frac{1}{5} :\phi_1 \rho_1: -\frac{2}{5} :\phi_2\rho_2: -\frac{2}{5} :\gamma^1\gammab_1: -\frac{4}{5} :\gamma^2\gammab_2:~,
\end{align}
The charges are fixed to be
\begin{align}
q_1 & = \frac{1}{5}~,&
q_2 & = \frac{2}{5}~,&
Q_1 & = -\frac{2}{5}~,&
Q_2 & = -\frac{4}{5}~,
\end{align}
so that
\begin{align}
c =\cb = r = \frac{3}{5}~,
\end{align}
while the charge of the unicorn operator $\cU = \gamma^1\gamma^2$ is $\qb_{\cU} = 4/5$.

There is no mystery here:  a second look at the ideal shows that it is equivalent to a more familiar one:
\begin{align}
\WW' = \la \phi_1^2, \phi_2^2\ra~,
\end{align}
i.e. a product of two decoupled $A_1$ minimal models, and we expect that in the IR the theory will simply flow to the decoupled product with central charge $\cb = 2/3 > 3/5$.  

This expectation can be verified by studying $\cV_{\tir}$.  We find a two-dimensional space of abelian currents in $H$, spanned by linear combinations of
\begin{align}
J^{1} & = -\ff{1}{3} :\phi_1 \rho_1: - \ff{2}{3} :(\gamma^1 + \phi_2 \gamma^2) \gammab_1:~,  \nonumber\\
J^{2} & = -\ff{1}{3} :\phi_2 \rho_2: - \ff{2}{3} :\gamma^2 (\gammab_2 - \ff{1}{2} \phi_2 \gammab_1):~.
\end{align}
This makes the accidental symmetry manifest.   Note that $J_{\tir}^{\text{naive}} = \ff{3}{5} J^1 + \ff{6}{5} J^2$.  As expected, these currents generate a nonlinear invariance of $\cW$ under  the
transformations
\begin{align}
\phi_1 &\to \phi_1  + \ff{1}{3}\epsilon_1\phi_1\quad &&\gamma^1
  \to\gamma^1 -  \ff{2}{3}\epsilon_1\left(\gamma^1 +
    \phi_2\gamma^2\right) + \ff{1}{3}\epsilon_2\phi_2\gamma^2\nonumber\\
\phi_1 &\to \phi_2  + \ff{1}{3}\epsilon_2\phi_2\quad &&\gamma^2\to
                                                               \gamma^2
                                                               -
                                                               \ff{2}{3}\epsilon_2\gamma^2~.
\end{align}
The OPEs of these currents yield the matrix $M = \frac13 \iden$.   The
coefficients $\tau_m$ of the distinguished linear combination
$J_{\text{ir}}$ can be readily calculated to be $(\frac13,\frac13)$,
leading to the expected charge  assignments and the expected
central charge of a product of minimal models.   Indeed, in $H^{2,0,0}$ we
can find two commuting spin--$2$ currents $T^{1,2}$, so that
the OPEs of $T^1, J^1$ and $T^2, J^2$ generate two commuting KMV algebras.
This is consistent with the expectation that the IR theory factors into a product.

In fact, we can do better by picking a different basis for the free fields generating $H$:
\begin{align}
\phi_1~, && \xi_1  &:= \rho_1~,&& \lambda^1 := \gamma^1 +\phi_2\gamma^2~,&&\lambdab_1  := \gammab_1~, \nonumber\\
\phi_2~, && \xi_2 &:= \rho_2+\gamma_2\gammab_1~,&& \lambda^2 := \gamma^2~,&&\lambdab_2 := \gammab_2 -\phi_2\gammab_1~.
\end{align}
The fields in the first line have trivial OPEs with fields in the second line, and the fields in each line have the OPEs of a standard free multiplet, i.e.~(\ref{eq:OPE}).  In terms of these fields the currents are
\begin{align}
J^{1} & = -\ff{1}{3} :\phi_1 \xi_1: -\ff{2}{3} : \lambda^1 \lambdab_1:~,&
J^{2} & = -\ff{1}{3} :\phi_2 \xi_2: -\ff{2}{3} :\lambda^2 \lambdab_2:~.
\end{align}
Moreover, the supercharge is given by
\begin{align}
\bQb & = -\oint \frac{dz}{2\pi i} \left( \gamma_1 W_1 + \gamma_2 W_2\right)  = -\oint \frac{dz}{2\pi i} \left (\lambda_1 \phi_1^2 + \lambda_2 \phi_2^2 \right)~ = \bQb_1 + \bQb_2~,
\end{align}
so that $H$ is isomorphic to the chiral algebra of the product of minimal models.  The change of variables is just a special instance of a holomorphic field redefinition in a LG theory.

The example also illustrates a right-moving enhancement:  the diagonal supercurrent multiplet which has $\bQ$ and $\bQb$ as the $\theta$ and $\thetab$ components is a sum of two commuting supercurrent multiplets of the two minimal models.  In this case there is also left-moving supersymmetry, and it has a beautiful realization in $H$~\cite{Witten:1993jg}.

\subsection{An $N=3$, $n=2$ example}
Next, we reconsider a  more involved example of a family of RG flows with accidental symmetries studied in~\cite{Bertolini:2014ela}.  In that work we used holomorphic field redefinitions to uncover the accidental symmetries.  We will now obtain the same results based solely on the structure of spin--$1$ currents in $H$.

\subsubsection*{The Model}

 The model consists of 3 fermi fields $\Gamma^A$, two
chiral fields $\Phi^i$, and a superpotential 
\begin{align}
\label{eq:devilmodel}
\cW_0 = \Gamma^T\alpha X\ ,
\end{align}
where we denote
\begin{align}
\Gamma= 
\begin{pmatrix} \Gamma^1 \\ \Gamma^2 \\ \Gamma^3 \end{pmatrix} 
\quad 
X  = \begin{pmatrix} \Phi_1^6 \\ \Phi_2^2 \\ \Phi_1^3 \Phi_2 \end{pmatrix}~,
\end{align}
and $\alpha_A^I$ is a $3\times3$ matrix of constants.  For generic
values of these, the potential preserves a unique $\GUL$ symmetry, and
normalizing the charges leads to $r=2$, $\cb = 3$, and charge
assignments
\begin{align}
\label{eq:devilcharges}
\xymatrix@R=0.0mm@C=1mm{~ & \Phi_1	& \Phi_2	&\Gamma^{1,2,3} \\
q	& \ff{1}{7} & \ff{3}{7}	& -\ff{6}{7}	  \\
\bqb	& \ff{1}{7} & \ff{3}{7}	& \ff{1}{7}
}
\end{align}

\subsubsection*{The IR symmetry algebra}

Implementing our program to construct $\cV_{\tir}^\circ$, we parameterize the general
$\GU(1)_{\text{uv}}\times\GU(1)_{R,{\text{uv}}}$--neutral spin--$1$  current as in
\eqref{eq:genspin1}, where 
\begin{align}
A^1 & = m_1\phi_1~, &A^2  &= m_2\phi_2 + m_3\phi_1^3~,
\end{align}
are the most general polynomials of the appropriate degrees, and $B$ is a $3\times 3$ matrix of constants.  Using~(\ref{eq:Jclosed}),
we find that $J$ is closed if and only if
\begin{align}\label{eq:jjclosed}
\alpha A = B\alpha~,
\end{align}
where 
\begin{align}
A &= \sum_{i=1}^3 m_i M_i~,
\end{align}
and
\begin{align}
M_1 &= \begin{pmatrix}
6 & 0 & 0 \\
0 & 0 & 0 \\
0 & 0 & 3
\end{pmatrix}~,
&
M_2 &= \begin{pmatrix}
0 & 0 & 0 \\
0 & 2 & 0 \\
0 & 0 & 1
\end{pmatrix}~,
&
M_3 &= \begin{pmatrix}
0 & 0 & 0 \\
0 & 0 & 2 \\
1 & 0 & 0
\end{pmatrix}~.
\end{align}
For invertible $\alpha$ we can solve \eqref{eq:jjclosed}
uniquely for $B$ given $A$ by 
\begin{align}\label{eq:invertible}
B  &= \alpha A\alpha^{-1}\ .
\end{align}
We thus have a three-dimensional space of spin--$1$, $\bQb$-closed fields
spanned by
\begin{align}
J^1 &=  -:\phi_1\rho_1: -  :\gamma^T B_1\gammab: ~,  \nonumber\\
J^2 &=  -:\phi_2\rho_2: -    :\gamma^T B_2\gammab:~, \nonumber\\
J^3  &= -\phi_1^3\rho_2 ~-  : \gamma^T B_3\gammab:~~,
\end{align}
with $B_i = \alpha A_i\alpha^{-1}$.   Computing the OPE of these, we
find that
\begin{align}
\Mh =\begin{pmatrix}
44 & 3 & 0\\
3 & 4  & 0\\
0 & 0 & 0
\end{pmatrix}~.
\end{align}
The kernel of $\Mh$ is spanned by $J^3$, which thus cannot be a
holomorphic current in the unitary IR theory.   Projecting this out,
we have a $\GU(1)^2$  KM algebra spanned by $J^{1,2}$, and calculating
$\tau$ as in~\eqref{eq:pickr}, we  find that the charges under
$(J_{\text{ir}},\Jb_{\tir})$ are given by 
\begin{align}\label{eq:devili}
\xymatrix@R=0.0mm@C=5mm{
~ 	& \theta		&\Phi_1		&\Phi_2		&\Gamma^{'1} 	&\Gamma^{'2}	&\Gamma^{'3} 	&&\\
q_{\tir}	& 0			& \ff{26}{167}	&\ff{64}{167}	&-\ff{156}{167}	&-\ff{128}{167}	&-\ff{142}{167} && \cb = 3\left( 1 + \ff{2}{167} \right)\\
\qb_{\tir}	& 1			&\ff{26}{167}	&\ff{64}{167}	&\ff{11}{167}	&\ff{39}{167}	&\ff{25}{167} 	&&
}
\end{align}
Here the $\Gamma^{'I}$ are the redefined fermi multiplets $\Gamma^{'I} = \sum_J \Gamma^J\alpha_{JI}$.
We can check that $J^3$ is not $\bQb$-exact: it is
associated with a chiral field, but \eqref{eq:devili} shows that it is
not holomorphic.

Next, we consider the low energy theory when $\alpha$ is not invertible.   If
$\alpha$ has rank 1 or less, the theory is singular (the ideal fails to be zero-dimensional), so the
remaining case is $\alpha$ of rank 2.  The theory then has a
free fermi field, and it is convenient to choose our basis for
$\gamma$ so  that this is $\gamma^3$.  In terms of the chiral algebra
this is the statement that $:\gamma^3\gammab_3:$ is a holomorphic
current with an associated holomorphic spin--$2$ current.  We will ignore this free field in what follows.
The remaining $N=n=2$ theory is
determined by a $2\times  3$ matrix,  which we continue to  denote by
$\alpha$.   The  construction of $H^{1,0,0}$ proceeds in much the same
way, with the fermi indices now taking only two values.  Equation \eqref{eq:jjclosed} now gives six linear equations in the seven
variables $m,B$.  This has a solution for generic (rank 2)  $\alpha$,
determining 
\begin{align}
J^1 = -:\phi_1\rho_1:-3:\phi_2\rho_2: - 6:\gamma^1\gammab_1:
  -6:\gamma^2\gammab_2:~. 
\end{align}
Normalizing this, and choosing a diagonal basis for the fermi fields, we find the charges and weights 
\begin{align}
\xymatrix@R=0.0mm@C=5mm{
~ 	& \theta		&\Phi_1		&\Phi_2		&\Gamma^{'1} 	&\Gamma^{'2}	&&	\\
q_{\tir}	& 0			& \ff{4}{31}	&\ff{12}{31}	&-\ff{24}{31}	&-\ff{24}{31}	&&	\cb = 3 \left( 1 + \ff{1}{31} \right)~. \\
\qb_{\tir}	& 1			&\ff{4}{31}		&\ff{12}{31}	&\ff{7}{31}		&\ff{7}{31}		&&	 
}
\end{align}

Finally, we can ask when there is an additional solution to
\eqref{eq:jjclosed} for a rank-2 matrix  $\alpha$.  Any non-singular theory requires $\alpha_{11} \neq 0$, and using that to simplify the algebra we find that an additional solution requires the vanishing of 
\begin{align}
D =\left(\alpha_{12}\alpha_{23} -\alpha_{13}\alpha_{22}\right)\Delta~,
\end{align}
where 
\begin{align*}
\Delta &= \alpha_{11}^2\alpha_{22}^2 -
  2\alpha_{11}\alpha_{12}\alpha_{21}\alpha_{22}  +
  \alpha_{11}\alpha_{12}\alpha_{23}^2 -
  \alpha_{11}\alpha_{13}\alpha_{22}\alpha_{23}  \nonumber\\
  &\qquad
  +\alpha_{12}^2\alpha_{21}^2  -
  \alpha_{12}\alpha_{13}\alpha_{21}\alpha_{23} + \alpha_{13}^2\alpha_{21}\alpha_{22}
\end{align*}
is the discriminant whose vanishing implies $\WW$ is not
zero-dimensional.  When the first factor in $D$ vanishes,  there is an
additional spin--$1$ class in $H^{1,0,0}$ represented by
\begin{align}
J^{2} &= -\ff{2}{3}(\alpha_{11}\alpha_{22}-\alpha_{12}\alpha_{21}):\phi_1\rho_1: 
+ (\alpha_{11}\alpha_{23}-\alpha_{13}\alpha_{21})\phi_1^3\rho_2 \nonumber\\
&\qquad+ (\alpha_{13}\alpha_{23} - 4\alpha_{12}\alpha_{21}):\gamma^1\gammab_1:
- (\alpha_{13}^2 - 4\alpha_{11}\alpha_{12}) :\gamma^1\gammab_2:\\ 
&\qquad+ (\alpha_{23}^2 - 4\alpha_{21}\alpha_{22}) :\gamma^2\gammab_1:
- (\alpha_{13}\alpha_{23} - 4\alpha_{12}\alpha_{21}):\gamma^2\gammab_2:  ~.\nonumber
\end{align}
Solving for $\tau_{1,2}$ as before, we find $J_{\tir}$ as a linear combination of these currents
and determine the charges to be (once again we choose a diagonal basis for the fermi fields)
\begin{align*}
\xymatrix@R=0.0mm@C=5mm{
~ 	& \theta		&\Phi_1		&\Phi_2		&\Gamma^{'1} 	&\Gamma^{'2}	&&	\\
q_{\tir}	& 0			&\ff{1}{7}		&\ff{1}{3}		&-\ff{6}{7}		&-\ff{2}{3}	 	&& \cb = 3\left(1+ \ff{1}{21}\right). \\
\qb_{\tir}	& 1			&\ff{1}{7}		&\ff{1}{3}		&\ff{1}{7}		&\ff{1}{3}	 	&&
}
\end{align*}
In fact, in this case there is also an additional holomorphic spin--$2$
current; the IR theory further factors into a product  of  two (2,2)
minimal models (and a free fermi field).  

We see that a perturbation of the last (most degenerate) case can be
chosen to flow to the previous (generic free fermion) case, from which
a perturbation takes us to the generic case.  For no $\alpha$ do we in
fact find $\cb=3$.  Thus, we see that the results obtained in~\cite{Bertolini:2014ela} are
consistent with the structure of the chiral algebra.

\section{Evidence for non-smooth $\mu \to 0$ limits} \label{s:terrible}
As we have seen, when the IR dynamics is determined by a compact, unitary SCFT, the
chiral algebra provides a powerful tool to partially characterize the
properties of this  SCFT.   If the chiral algebra is not consistent
with the constraints of unitarity, the  assumptions underlying the
calculations above must be re-examined.  Here we present two examples of this phenomenon.
In the first example we will see that, while the chiral algebra is incompatible with a compact unitary SCFT, we nevertheless have a plausible description of the IR fixed points.  We have not been able to give such a description in our second example.

\subsection{A moduli stack with bad orbits} \label{ss:stack}

We discussed in the introduction the topological quotient
\begin{align}
\cM = \left\{ \cM_{\tuv} \setminus \Delta_0 \right\} / \cG~
\end{align}
as a description of the IR fixed points and pointed out three subtle aspects of such a description:  the quotient is not a separated space, and the quotient topology fails to separate orbits that correspond to distinct IR physics; there can be additional IR equivalences: two distinct orbits can flow to the same IR fixed point; finally, $\Delta_0$ may not be full discriminant of the theory.  The example we discuss next will illustrate the first two issues, and the second example may be an example of the second feature.

Consider a theory with $N=n=2$ and ideal generators
\begin{align}
W_1 & = \phi_1^4 + a \phi_1 \phi_2^2~,&
W_2 & = \phi_2^5 + b \phi_1^6 \phi_2 + c \phi_1^3 \phi_2^3~.
\end{align}
This is the most general ideal consistent with the given leading monomials $\phi_1^4 \subset W_1$, $\phi_2^5 \subset W_2$ and preserving a quasi-homogeneity relation with $3q_1 = 2q_2$.   So, for this family $\cM_{\tuv} = \C^3$, with the three affine coordinates $(a,b,c)$.  The ideal is zero-dimensional when the combination of parameters $z = a (c-ab) \neq 1$, i.e. $\Delta_0 = \{ z=1\}$.  

To study the quotient, we consider the most general holomorphic field redefinitions consistent with the UV symmetries.  A convenient presentation is given in terms of three parameters, $\alpha_1,\alpha_2\in \C^\ast$ and $\beta \in \C$:
\begin{align}
\phi_1 &\to \alpha_1 \phi_1~,&
\phi_2 &\to \alpha_2 \phi_2~,&
\gamma^1 &\to \alpha_1^{-4} (\gamma^1 + \beta \alpha_1^6 \alpha_2^{-4} \phi_1^2 \phi_2 \gamma^2)~. &
\gamma^2 &\to \alpha_2^{-5}\gamma^2~,
\end{align}
Under this action the parameters $(a,b,c)$ transform to 
\begin{align}
(a,b,c) \mapsto (\zeta^{-1} a, \zeta^2 (b+\beta), \zeta (c +  \beta a))~,
\end{align}
where $\zeta = \alpha_1^{3} \alpha_2^{-2}$. 
The group of field redefinitions that acts on the parameter space is therefore  $\cG = \C^\ast \rtimes \C$, with elements $(\zeta,\beta)$ and group multiplication given by
\begin{align}
 (\zeta_2,\beta_2) \cdot (\zeta_1 ,\beta_1) = (\zeta_2 \zeta_1, \beta_2 + \zeta_2^{-1} \beta_1)~.
\end{align}
We can choose the parameter $\beta$ to set $b=0$, and with this partial fixing we find
\begin{align}
\cM = \left\{ \C^2 \setminus \{ ac =1 \} \right\} / \C^\ast~,
\end{align}
where the $\C^\ast$ action is
\begin{align} 
(a,c) \mapsto (\zeta^{-1} a, \zeta c)~.
\end{align}
$\cM$ is a non-separated space, and it is important to keep track of its stack structure.   Generic $\C^\ast$-orbits are labeled by the non-zero value of the invariant $z = ac$, but $z = 0$ corresponds to three distinct orbits:
\begin{align}
O_{(i)} &= \{ \text{ $a=0$, $c=0$ } \};   \nonumber\\
O_{(ii)} & = \{\text{$a \neq 0$, $c = 0$ } \}; \nonumber\\
O_{(iii)} & = \{\text{ $a=0$, $c\neq 0$} \}.
\end{align}
The first orbit has dimension zero and corresponds to a product of two minimal models.  The second orbit has dimension one and also has a simple interpretation from the point of view of the minimal model product: turning on $a \neq 0$ can be interpreted as adding a relevant deformation to the minimal model product that breaks the UV symmetry $\GU(1)^2 \to \GU(1)$, and it is not surprising that the resulting theory should have a lower central charge and smaller KMV algebra.  These expectations are verified by studying the chiral algebra for these values of the parameters.  

The last orbit is the most surprising one.  Using the minimal model R-charge assignments, the operator corresponding to $c\neq 0$ is clearly irrelevant, so we expect to recover the minimal model product in the IR.\footnote{This interpretation is consistent with the conjectural description of the IR given in~\cite{Bertolini:2014ela}.} And thus, if the IR limit $\mu \to 0$ were smooth, $H$ should be isomorphic to the chiral algebra of the minimal model product.  
Studying the chiral algebra in detail shows that this is not the case for parameters belonging to $O_{(iii)}$: we find $\cV_{\tuv} = \cV_{\tir}$, and the central charge is identical to that of $O_{(ii)}$.  It is also easy to see that the chiral algebra obtained from $O_{(iii)}$ is not isomorphic to that of $O_{(ii)}$.\footnote{For instance, for either orbit $H^{0,2q_1,2q_1}$ is one-dimensional and generated by $\phi_1^2$, but while on $O_{(iii)}$ the operator $\phi_1^2$ is nilpotent, it is not nilpotent on $O_{(ii)}$.}

The appearance of $O_{(iii)}$ raises a number of puzzles.  For example, we expect that there is a family of SCFTs labeled by the parameter $z$, and $z=0$ appears to be a smooth point in that family.  If both $O_{(ii)}$ and $O_{(iii)}$ belong to $z=0$, then it appears that the $z=0$ limit in the SCFT moduli space is ambiguous.

As we will now argue, the chiral algebra for $O_{(iii)}$ is not consistent with unitarity constraints of the SCFT.  Therefore, the IR physics must be different from the naive description based on the chiral algebra.  A plausible explanation is that theories belonging to $O_{(iii)}$, like those belonging to $O_{(i)}$, flow to the minimal model product SCFT, while the theories parameterized by $z\neq 1$, including the orbit $O_{(ii)}$ at $z=0$, flow to a family of SCFTs that realize the structure of the UV chiral algebra.

\subsubsection*{Non-unitarity for the $O_{(iii)}$ orbit}
Working at generic $z$ and using the unique symmetry in $\cV_{\tir}^\circ$, we obtain the following charge assignments for the fields:
\begin{align}
\xymatrix@R=0.0mm@C=5mm{
~ 	& \theta		&\Phi_1		&\Phi_2		&\Gamma^1 	&\Gamma^2	&&	\\
q_{\tir}	& 0			& \ff{3}{23}	&\ff{9}{46}	&-\ff{12}{23}	&-\ff{45}{46}	&&	\frac{\cb}{3} =  \frac{27}{23}~. \\
\qb_{\tir}	& 1			&\ff{3}{23}		&\ff{9}{46}	&\ff{11}{23}		&\ff{1}{46}		&&	 
}
\end{align}
The unicorn operator has charge $\qb_{\cU} = 1/2 < \cb/3$.  However, we can look at the unitarity bounds for a more general class of operators $\cU_{m_1,m_2} = \phi_1^{m_1} \phi_2^{m_2} \gamma^1\gamma^2$.  

The quotient ring $R_{\WW}$ has dimension $20$, and a basis can be taken to be
\begin{align}
\xymatrix@R=0.0mm@C=5mm{
1 					& \phi_1~,						& \phi_1^2~,					&\phi_1^3~,		 \\
\phi_2~,				& \phi_1 \phi_2~,				& \phi_1^2\phi_2~,				&\phi_1^3\phi_2~, \\
\phi_2^2~,				& \phi_1 \phi_2^2~,				& \phi_1^2\phi_2^2~,				&\underline{\phi_1^3\phi_2^2}~, \\
\phi_2^3~,				& \underline{\phi_1 \phi_2^3}~,		&\underline{\phi_1^2\phi_2^3}~,		&\underline{\phi_1^3\phi_2^3}~, \\
\underline{\phi_2^4}~,		& \underline{\phi_1 \phi_2^4}~,		& \underline{\phi_1^2\phi_2^4}~,		&\underline{\phi_1^3\phi_2^4}~, \\
}
\end{align}
where we underlined the monomials that can potentially lead to $\cU_{m_1,m_2}$ that violates the unitarity bound $\qb_{\tir} \ge \cb/3$.  Every $\cU_{m_1,m_2}$ is $\bQb$-closed, but some of these operators are in the image of $\bQb$.  A $\bQb$-exact $\cU_{m_1,m_2}$ must be of the form $\bQb \cdot X$, with 
\begin{align}
X = \sum_{a,b=1}^2 \gamma^a f^{ab}(\phi) \rho_b + \gamma^1\gamma^2 \sum_{a=1}^2 g^a(\phi)\gammab_a~.
\end{align}
We find
\begin{align}
\bQb\cdot X = \gamma^1\gamma^2\left(f^{11} W_{2,1} + f^{12} W_{2,2} - f^{21} W_{1,1} - f^{22} W_{1,2} - g_1 W_1 - g_2 W_2\right)~.
\end{align}
Working with the ideal at $b =0$, i.e. with
\begin{align}
W_1 & = \phi_1^4 + a \phi_1 \phi_2^2~,&
W_2 & = \phi_2^5 + c \phi_1^3 \phi_2^3~,
\end{align}
we can see that when $a \neq 0$ all potential $\cU_{m_1,m_2}$ that could violate the unitarity bound are $\bQb$-exact.  In detail, taking
$f^{22}$ and $f^{12}$ to be the only non-zero components, we have
\begin{align}
\bQb\cdot X & =\gamma^1\gamma^2\left( f^{12} (5\phi_2^4 + 3c \phi_1^3 \phi_2^2) -2a f^{22} \phi_1 \phi_2\right)~.
\end{align}
If we set $f^{12} = 0$, and $f^{22} = \phi_1^{m_1-1} \phi_2^{m_2}$, we eliminate all underlined $\cU_{m_1,m_2}$ with $m_1>0$ from cohomology.
To show that $\cU_{4,0}$ is also $\bQb$-exact, we take $f^{12} = 1/5$ and $f^{22} = \ff{3c}{10a} \phi_1^2\phi_2$.

On the other hand, when $a = 0$ we find
\begin{align}
\bQb\cdot X  & = \gamma^1\gamma^2\left( 3cf^{11}\phi_1^2\phi_2^3 + f^{12} (5\phi_2^4 + 3 c\phi_1^3\phi_2^2) -4 \phi_1^3 f^{21} -g_1 W_1 -g_2 W_2\right)~, 
\end{align}
and we see that there is no way to obtain $\cU_{1,3} = \phi_1\phi_2^3\gamma^1\gamma^2$ in the image of $\bQb$.

\subsection{A rigid example with a unitarity-violating unicorn} \label{ss:badunicorn}
In the previous example we saw the orbit $O_{(iii)}$ for which the chiral algebra was not consistent with a unitary compact SCFT.  However, we had a natural guess for the IR fixed point:  the minimal model product that could be realized in the same $\cM_{\tuv}$ as any point in $O_{(iii)}$. We now present an example where such a guess is unavailable because $\cM$ consists of a single orbit.

We consider a family of models with $N=n=3$ and  the superpotential interaction 
\begin{align}\label{eq:badunicorn}
\cW =\Gamma^1\Phi_1\Phi_2 + \Gamma^2\left(\Phi_2^3 +
  \Phi_1\Phi_3\right) + \Gamma^3\left(\Phi_3^2 + \Phi_1^p\right)~
\end{align}
labeled by the integer $p\ge 2$. 
These theories have $k=1$,  and if we assume that $\cV_{\tir} = \cV_{\tuv}$, the charges are given by
\begin{align}\label{eq:baducharges}
\xymatrix@R=0.0mm@C=5mm{
~ 	& \theta		&\Phi_1		&\Phi_2		&\Phi_3		&\Gamma^1 	&\Gamma^2	&\Gamma^3	&&	\\
q_{\tir}	&0 &6\sigma   &(p+2)\sigma &3p\sigma &-(p+8)\sigma &-3(p+2)\sigma &-6p\sigma && \\
\qb_{\tir}	& 1			&6\sigma	&(p+2)\sigma	&3p\sigma	&1-(p+8)\sigma	&1-3(p+2)\sigma	 &1-6p\sigma			&&
}
\end{align}
where 
\begin{align}
\sigma = \frac{p+1}{2(3p^2 +4p+5)}~.
\end{align}
 The central charge predicted by these is
\begin{align}
\label{eq:badcbar}
\frac{\cb}{3} = 6(p+1)\sigma~.
\end{align} 
The
unicorn $\cU$ has charge
\begin{align}
\label{eq:badunicorncharge}
\qb_\cU = \frac{8(p^2+2)}{p+1} \sigma~, 
\end{align}
and $\qb_{\cU} > \cb/3$ for $p > 5$.  

To investigate this further, we will now assume $p+2 \ge 7$ is prime.\footnote{We suspect our results hold more generally, but since this technical assumption already yields an infinite family of theories, we will stick to this subclass.}
We will show that, unlike our previous examples with an apparent unitarity violation, the cohomology group $H^{1,0,0}$ is one-dimensional:  $\cV_{\tir} = \cV_{\tuv}$.  Assuming that the theory flows at low energies to a compact SCFT leads to a contradiction: the chiral algebra constructed in the UV using free field OPEs cannot
possibly be the chiral algebra of a unitary compact SCFT. 

The restriction that $p+2$ is prime guarantees that the only holomorphic redefinitions consistent with quasi-homogeneity are re-scalings of the chiral fields: $\Phi_a \to \alpha_a \Phi_a$ and $\Gamma^A \to \beta^A \Gamma^A$.  Moreover, the potential $\cW$ is rigid: the monomials shown are the only ones that are consistent with quasi-homogeneity, and all of them must be present in order for the ideal to be zero-dimensional.  Any zero-dimensional ideal consistent with this quasi-homogeneity relation can be brought to the canonical form in~(\ref{eq:badunicorn}) by a holomorphic field redefinition.

The claims of the previous paragraph follow by straightforward numerological considerations.  We will not give all of these manipulations here but just list two to give the reader the flavor of the simple manipulations involved.  Consider, for example, a monomial $\phi_1^{n_1} \phi_2^{n_2} \phi_3^{n_3}$ that can show up in a field redefinition of $\phi_3$:  \begin{align}
\phi'_3 = \alpha_1 \phi_3 + \alpha_2 \phi_1^{n_1} \phi_2^{n_2} \phi_3^{n_3}+\cdots~.
\end{align}
 Quasi-homogeneity requires $n_3 =0$ and 
\begin{align}
6 n_1 + (p+2) n_2  = 3p~.
\end{align}
We rewrite this as
\begin{align}
(3-n_2) (p+2) = 6 (n_1 + 1)~
\end{align}
so that $n_2 \in \{0,1,2\}$.  But, each of these possibilities is inconsistent with $p+2$ being prime.

As a second example, consider a field redefinition of the form  
\begin{align}
\gamma'^{1} = \beta_1 \gamma^1 +\beta_2\phi_1^{n_1} \phi_2^{n_2} \phi_3^{n_3} \gamma^3+\cdots~.
\end{align}
This requires
\begin{align}
6p - (p+8) = 6 n_1 + (p+2) n_2 + 3p n_3~,
\end{align}
which we rearrange as
\begin{align}
(5-n_2-3n_3) (p+2) = 6 (n_1+3-n_3)~.
\end{align}
That is only possible if $n_1= n_3-3+ l (p+2)$ for some integer $l$.  Plugging that into the relation leads to $n_2 =5-3n_3-6l$, which can only be satisfied if $l \le 0$.  But now, combining our expressions for $n_1$ and $n_2$, we have $3n_1 + n_2 = 3lp-4  \le 0$, which is a contradiction.

Following in the same vein, we establish all of the assertions for the potential and redefinitions in this class of theories.

%

 This restricts $J$ to the simple form
 \begin{align}
   J = \sum_{a=1}^3 A_a:\phi_a\rho_a: + \sum_{A=1}^3 B_A:\gamma^A\gammab_A:
 \end{align}
 with some constants $A_a$ and $B_A$ associated to linear rotations of the chiral fields.   Up to an overall normalization, the only
$\bQb$-closed  combination is then given by taking the coefficients to be the charges $q_{\tir}$ in~(\ref{eq:baducharges}).

If the theory with $p=11$, for example, flows to a unitary compact SCFT, then the unicorn operator $\cU \in H$ must correspond to a chiral primary field of spin $3/2$, and there must be an operator $T_{\tir}$ in $H$ that represents the left-moving energy-momentum tensor of the SCFT.  We showed that $T_{\tir}$ exists and is the unique spin---$2$ operator consistent with the KMV algebra $\cV_{\tir}$.  Thus, the R-charge of $\cU$ $\qb_{\cU}$ is determined by~(\ref{eq:qbarfromspin}), while $\cb = c$ is determined from the $T_{\tir}T_{\tir}$ OPE; the two take the values quoted in~(\ref{eq:badunicorncharge},\ref{eq:badcbar}), and $\qb_{\cU} > \cb/3$.  So, irrespective of any possible non-abelian enhancements or extra right-moving spin---$1$ currents, we see that for $p=11$ the chiral algebra of the model is inconsistent with a compact unitary SCFT.

The case $p=5$ is marginal, in the sense that $\qb_\cU = \cb/3$.
This does not explicitly violate the unicorn bound.  As discussed
above, however, a field with this charge is related by spectral flow
to a holomorphic field.  In the case at hand, this predicts a holomorphic
spin--$9/2$ field of charges  $(q_{\tir},\qb_{\tir}) = (-3,0)$.  We can indeed construct such  a
field.   Since this  is holomorphic, a unitary theory requires  the
existence of a conjugate spin--$9/2$ field of charges $(3,0)$.   An
explicit calculation shows no cohomology exists at these charges,
showing that the marginal $p=5$ case also cannot flow to a unitary, compact
SCFT.  

Before speculating on how our model fails to flow to a compact,
unitary SCFT, it is wise to check our technical assumptions.  We
assumed, for example, that for $k=1$ the left-moving weights are
determined by $T_{\tir}$ with $\tau$  given by \eqref{eq:pickr}.   If
there are other spin--$2$ fields in cohomology, it is conceivable that we
have misidentified which of them represents $T$.  To check this, we
explicitly constructed $H^{2,0,0}$ for $p=11$ and for $p=5$. We verified
that in each case this is three-dimensional and is spanned by
$T_{\tir}$, $\p J$, and $:JJ:$ as expected.  
$T_{\tir}$ is the unique combination that combines with $J$ to a KMV
algebra  $\cV_{\tir}$.

\section{Discussion} \label{s:discussion}
In this work we studied the chiral algebra of (0,2) LG theories.  We obtained a number of results under the assumption that a LG theory specified by a zero-dimensional quasi-homogeneous ideal flows to a compact unitary SCFT in the IR.  In particular, we found a Kac-Moody-Virasoro subalgebra $\cV_{\tir}$ of the chiral algebra and a representative of the holomorphic energy momentum tensor $T_{\tir}$.  When the $\mu\to0$ limit of the chiral algebra is smooth, so that this structure matches the chiral algebra of the IR SCFT, these results give us a great deal of information about the fixed point theory.

In a number of examples we saw that all of this machinery is consistent with expectations and constraints that arise from unitarity of the SCFT.  Nevertheless, we also obtained a troubling result: there are models in our class that cannot flow to compact SCFTs that realize this chiral algebra structure.  Our understanding of this is rather crude:  we studied special operators $\cU \in H^{N/2}$ and the consequences of the bound $\qb_{\cU} \le \cb/3$.  We found examples where the chiral algebra predicts $\qb_{\cU} > \cb/3$.

Our results used the assumptions of compactness, supersymmetry, and unitarity of the IR theory.  Since our UV theory is defined by an asymptotically free Lagrangian, we expect the IR limit to be unitary.  It seems unlikely that supersymmetry is broken in this class of theories: there is an $\cR$ multiplet along the flow, and spontaneous symmetry breaking would be inconsistent with the non-zero elliptic genus.  On the other hand, compactness can perhaps be violated through the details of the D-terms, over which we have no control.  If the IR theory is not compact, then it is no longer true that symmetries must be realized by Kac-Moody currents, and most of our results would no longer hold.

We can understand this at the level of the chiral algebra as follows.  While the holomorphic renormalization scheme we assume will preserve the algebraic structure of $H$ along the RG flow, the normalization of the operators in $H$ is not similarly protected, since the two-point function inevitably involves the product of chiral and anti-chiral operators.  If we normalize the operators so that the two-point function takes a canonical form at scale $\mu$, it might be that some of the operators are scaled to zero as $\mu\to 0$, and the chiral algebra, when rewritten in terms of the rescaled operators, might take a different form as $\mu \to 0$.  For example, it is conceivable that some of the terms in $\cW$, when re-expressed in the normalized fields, are scaled to zero as $\mu \to 0$.  When this happens it is possible that either there are new symmetries that emerge in the IR (this would account for the puzzles of the example in section~\ref{ss:stack}), or the spin zero chiral ring becomes infinite-dimensional (this would signal non-compactness and account for the puzzle of the example in section~\ref{ss:badunicorn}).

In either case, the resolution of the unitarity puzzles seen in the chiral algebra seems to require tools that involve non-chiral quantities, whether it be control over the D-terms of the Lagrangian theory, or computation of the renormalized two-point functions.

We expect that in general $\qb_{\cU} \le \cb/3$ is just the first of many non-trivial unitarity conditions.  It is conceivable that the chiral algebra of a theory may satisfy all unitarity conditions for operators of some spin $s< S_{\text{min}}$ but exhibit a failure for some operator $\cO \in H^{S_{\text{min}}}$, and it is conceivable that $S_{\text{min}}$ grows with $N$ or $n$.  It should also be clear that the unitarity violations we have seen will be a fairly generic phenomenon.  For example, most $N=n=2$ theories consistent with our assumptions exhibit the puzzles of the example in section~\ref{ss:stack}:  there are orbits like $O_{(iii)}$ that correspond to seemingly irrelevant deformations of a known theory, yet the chiral algebra does not match the known theory.

It would be interesting to find a finite set of conditions on $\WW$ with fixed $n$ and $N$ that would guarantee a unitary spectrum, but even such a set of conditions would merely provide necessary conditions for the fixed point to be a compact unitary SCFT.  We are thus led to important questions: given a (0,2) LG theory specified by an ideal $\WW$, what are the necessary and sufficient conditions for this theory to flow to a compact unitary SCFT that realizes the LG chiral algebra?   if a (0,2) LG theory does not satisfy these conditions, what is its low energy limit?

These questions also apply to other seemingly well-behaved (0,2) Lagrangian theories such as the linear sigma models;
answering them is a prerequisite to extending the successes of seemingly similar (2,2) renormalization group flows to the (0,2) setting, but they require new tools.

\appendix

\section{Diagonal R-symmetry action} \label{app:Rsym}
Let $\WW \subset \C[\phi_1,\ldots,\phi_n]$ be a zero-dimensional quasi-homogeneous ideal with the property that no variable $\phi_a$ appears linearly in any generator of $\WW$.\footnote{The last assumption excludes any mass terms from our LG theory.}  
Suppose that the quotient ring $R_{\mathbb{W}} = \C[\phi_1,\ldots,\phi_n] / \mathbb{W}$ admits an R-symmetry action $\deltah$ which descends from an action $\delta$ on $\C[\phi_1,\ldots,\phi_n]$ 
\begin{align}
\delta \phi_a = i \alpha \zeta_a(\phi_1,\ldots,\phi_n) \in \C[\phi_1,\ldots,\phi_n]
\end{align}
with the following properties:
\begin{enumerate}
\item the action preserves the ideal:  if $g \in \mathbb{W}$, then $\delta g \in \mathbb{W}$;
\item the R-symmetry acts as a linear diagonalizable operator on $R_{\mathbb{W}}$;
\item the linear action is compatible with the product in the quotient ring;
\item there is a one-dimensional invariant subspace;
\item the remaining eigenvalues are positive;
\item the action preserves the unique vacuum:  $\zeta_a|_{\phi = 0} = 0$ for all $a$.
\end{enumerate}
With apologies to commutative algebra artisans, we will prove the following theorem: it is possible to find a polynomial change of variables
\begin{align}
\phi'_a = f_a(\phi_1,\ldots,\phi_n)
\end{align}
such that the R-symmetry has a diagonal action on the $\phi'_a$.

Before launching into the details of the proof, we will comment on why $\deltah$ should have such a lift $\delta$ and discuss the list of properties we expect of $\delta$.  If our theory flows to a compact SCFT, then the vector space underlying the quotient ring $R_{\WW}$ is isomorphic to the Hilbert space of spin zero chiral primary states in the SCFT.  The R-symmetry action preserves this subspace of the full Hilbert space and must be represented by a linear diagonalizable action $\deltah$ that has a one-dimensional invariant subspace corresponding to the unique $\SL(2,\C)$ vacuum.  Unitarity requires that the remaining eigenvalues of $\deltah$ be positive, and compatibility with the OPE requires $\deltah$ to act as a derivation with respect to multiplication in $R_{\WW}$.  This is enough to argue that $\deltah$ has a lift to an action $\delta$ on $\C[\phi_1,\ldots,\phi_n]$ with the first $5$ properties claimed above.  

The last property, that the lift preserves the unique vacuum configuration $\phi =0$, is an extra condition.  We will impose it since without it it seems difficult to provide a description of the IR physics in terms of the UV fields.  At any rate, such a description would require some information beyond the superpotential couplings encoded in the ideal $\WW$.

\subsection*{Proof of theorem}
We use the R-symmetry action to grade the vector space $R_{\WW}$:
\begin{align}
R_{\mathbb{W}}  = R^{0} \oplus R^{\qb_1} \oplus R^{\qb_2} \oplus \cdots \oplus R^{\qb_t}~,
\end{align}
with $R^{0} = \C$, and $0<\qb_1 <\qb_2< \cdots < \qb_t$.  Because the ideal is zero-dimensional, the set of $\{\qb_i\}$ is finite, and in particular there is a smallest non-zero charge $\qb_1$.  We also define the corresponding filtration by setting
\begin{align}
\label{eq:filter}
\cF^{\qb_i} = R^{\qb_i} \oplus R^{\qb_{i+1}}  \oplus \cdots \oplus R^{\qb_t}~,
\end{align}
so that $\cF^{\qb_{i+1}} \subset \cF^{\qb_i}$.  Multiplication in the ring respects this filtration:  if $[P_i] \in \cF^{\qb_i}$ and $[P_j] \in \cF^{\qb_j}$, then $[P_i P_j] \in \cF^{\qb_i + \qb_j}$~.  Here we used the notation that for any polynomial $P$, $[P]$ 
denotes its equivalence class in $R_{\WW}$.

Any polynomial $P$ can be written as 
\begin{align}
P = c + \sum_{a} v_a \phi_a + G~,
\end{align}
where $c$ and $v_a$ are constants, while $G$ is non-linear in the $\phi_a$.  The R-symmetry action  $\delta \phi_a = \zeta_a$ has a similar expansion, but our assumption 6 requires the constant term to be absent, so that
\begin{align}
\delta \phi_a = \zeta_a = \sum_{b} R_{ab} \phi_b + F_a~,
\end{align}
where $R_{ab}$ is a constant matrix, and $F_a$ is non-linear in the $\phi_1,\ldots,\phi_n$.  With this notation it is easy to see that if $[P] \in R^{\qb}$ with $\qb>0$, then it must be that $c = 0$, and the vector $v$ is an eigenvector of the matrix $R$.

Every variable $\phi_a$ can also be expanded according to the grading.  More precisely, we can find polynomials $\Phi_a^{\qb_i} \in \C[\phi_1,\ldots,\phi_n]$ with $[\Phi_a^{\qb_i}] \in R^{\qb_i}$, such that
\begin{align}
\label{eq:phidecomp}
\phi_a = C_a + \Phi_a^{\qb_1} + \Phi_a^{\qb_2} + \cdots + \Phi_a^{\qb_t} \mod \WW~.
\end{align}
The constants $C_a$ must be zero:  our ideal is quasi-homogeneous and zero-dimensional and has no linear terms in its generators, which means for every $a$  there is an integer $k>1$ so that $\phi_a^k \in \WW$.  

Since the $\Phi_a^{\qb_i}$ are R-symmetry eigenstates, by our previous statement it must be that
\begin{align}
\Phi_a^{\qb_i} = \sum_{b}  v_{ab}^i \phi_b + G^i_a~,
\end{align}
where the $G^i_a$ are non-linear, and the $v^i_{a}$ are eigenvectors of the matrix $R$ with eigenvalue $\qb_i$.  But now, coming back to~(\ref{eq:phidecomp}) and taking the linear terms, which must vanish by themselves, we find that for all $a$
\begin{align}
\phi_a = \sum_{b,i} v^i_{ab} \phi_b ~.
\end{align}
This can only hold if the eigenvectors $v^i_a$ of $R$ span $\C^n$, i.e. $R$ is diagonalizable.  We can then make a linear change of variables so that in the new variables the R-symmetry action is
\begin{align}
\delta \phi_a = \sigmab_a \phi_a + F_a~,
\end{align}
where the $\sigmab_a$ denote the eigenvalues of the matrix $R$.  Plugging this back into~(\ref{eq:phidecomp}) rewritten in the new variables, we find that $v^i_{ab} (\sigmab_b - \qb_i) = 0$, so that for $\qb_i \neq \sigmab_b$ we have $\Phi_a^{\qb_i} = G^i_a$, and
\begin{align}
\Phi_a^{\sigmab_a} = \phi_a - \sum_{i,\qb_i \neq \sigmab_a} G^i_a \mod \WW~.
\end{align}

\subsubsection*{Improving the variables}
For each $\phi_a$ let $r_a$ denote the R-charge such that $[\phi_a] \in \cF^{r_a}$ but $[\phi_a] \not\in \cF^{\qb}$ with $\qb>r_a$.  We can order the variables so that $r_1 \le r_2 \le \cdots \le r_n$.  Whenever
$\sigmab_a > r_a$,  the decomposition of $\phi_a$ has the form
\begin{align}
\phi_a = \Phi_a^{r_a} + \cdots + \Phi_a^{\sigmab_a} + \cdots + \Phi_a^{\qb_t}  \mod \WW~,
\end{align}
and each $\Phi_a^{\qb_i}$ with $\qb_i <\sigmab_a$ is non-linear in the variables.  Using the filtration structure, we see that we can choose a representative for $\Phi_a^{\qb_i}$ that only depends on variables $\phi_b$ with $r_b <\qb_i$.  We will now argue inductively that we can make a change of variables so that all such terms are absent, and $r_a = \sigmab_a$ for the new variables.   

The first step is to find the smallest $r_a$ for which $\sigmab_a > r_a$.  For each such variable
\begin{align}
\phi_a = F_a + \Phi_a^{\sigmab_a} + \cdots \mod \WW~,
\end{align}
where $F_a$ is a nonlinear function of the variables that only depends on variables $\phi_b$ with $r_b = \sigmab_b <r_a$.
For each such $\phi_a$ we make a change of variables
\begin{align}
\phi'_a = \phi_a - F_a~,
\end{align}
while leaving the $\phi_b$ with $r_b = \sigmab_b <r_a$ unchanged. This is clearly invertible, and the new variables $\phi'_a$ will satisfy $\sigmab'_a = r'_a$.  Repeating this procedure as necessary up to the maximum $r_a$, we obtain the desired result:  a new set of variables $\phi'_a \in \cF^{\sigma_a}$ with $\sigmab_1 \le \sigmab_2 \le \cdots \le \sigmab_n$.

When written in terms of the new variables it follows that
\begin{align}
\Phi_a^{\sigmab_a} = \phi'_a + G_a~,
\end{align}
and the filtration implies that the non-linear polynomial $G_a$ can only depend on variables $\phi'_b$ with $b<a$.  That means the change of variables
\begin{align}
\phi''_a = \phi'_a + G_a(\phi'_1,\ldots, \phi'_{a-1})
\end{align}
is invertible, and the new variables have R-symmetry transformation
\begin{align}
\delta \phi''_a = \sigmab_a \phi''_a \mod \WW~.
\end{align}
This is our desired result: there is a holomorphic change of variables $\phi_a'' = \phi_a''(\phi_1,\ldots,\phi_n)$ so that in cohomology the R-symmetry action can be taken to be $\delta \phi''_a = \sigmab_a \phi''_a$.

\subsection*{Extension to $\bQb$ cohomology at spin--$1/2$ for $N=n$}
Similar considerations apply to $H^{1/2}$, the $\bQb$ cohomology at spin $1/2$:  this space must admit a diagonalizable R-symmetry action.  We will show that for $N=n$ this implies that we can make a holomorphic field redefinition
\begin{align}
\gamma'^A = f_A^B(\phi) \gamma^B~,
\end{align}
so that $\gamma'^A$ are eigenstates of the R-symmetry action.

There are two sorts of terms that can potentially appear in $H^{1/2}$:   $\sum_A g_A(\phi)\gamma^A$ and  $\sum_A h^A(\phi) \gammab_A$.  Note that these must be separately $\bQb$-closed.   When $N=n$, a $\bQb$-closed operator of the second type must be $\bQb$-exact:  as we discuss in appendix~\ref{app:Koszul}, if this is not the case, then $\WW$ cannot be zero-dimensional.

Thus, $H^{1/2}$ is spanned by operators $\cO[f] = \sum_A f_A(\phi)\gamma^A$.  Modding out by $\bQb$-exact terms leads to an equivalence relation
\begin{align}
\label{eq:equivalence}
f_A \sim f_A + \sum_{b=1}^n W_{A,b} g_b (\phi) + \sum_{B=1}^N h_A^B(\phi) W_B~.
\end{align}
This ensures that $H^{1/2}$ is finite-dimensional when $\WW$ is zero-dimensional.

We now grade $H^{1/2}$ by the R-symmetry action as before:
\begin{align}
H^{1/2} = R^{\qb_1} \oplus R^{\qb_2} \oplus \cdots \oplus R^{\qb_t}~,
\end{align}
with $0\le \qb_1 <\qb_2< \cdots < \qb_t$.  Because the cohomology is finite-dimensional, the set of $\{\qb_i\}$ is finite, and there is a minimum non-zero charge $\qb_1$.  We also define the $\cF^{\qb_i}$ as in~(\ref{eq:filter}).

In fact, $\qb_1$ is positive.  Otherwise $H^{1/2}$ contains an R-neutral spin--$1/2$ operator $\Psi$, i.e. a holomorphic fermion.  In a unitary CFT this operator must have a holomorphic conjugate $\Psi^\dag$ with
\begin{align}
\Psi^\dag(z) \Psi(w) & \sim \frac{1}{z-w}~,
\end{align}
and $\Psi^\dag$ must also belong to the chiral algebra.  However, in our free field representation, the only possibility for such an operator is $\Psi^\dag = \sum_A h^A(\phi) \gammab_A$, and we already argued these to be absent when $\WW$ is zero-dimensional.  So, $\qb_1 >0$.

Since we assume that there are no linear terms in our ideal generators (i.e. no mass terms), each field $\gamma^A$ is non-trivial in $H^{1/2}$ and has an expansion in terms of R-symmetry eigenstates:
\begin{align}
\gamma^A = \gamma^{A, \qb_1} + \gamma^{A,\qb_2} + \cdots + \gamma^{A,\qb_t} \mod \WW~,
\end{align}
and the R-symmetry action has a lift $\delta$ with
\begin{align}
\delta \gamma^A = \sum_{B} ( R^A_B + M^A_B(\phi) ) \gamma^B~,
\end{align}
where $R$ is a constant $N\times N$ matrix, while $M^A_B$ depends on the $\phi_a$.  These statements are true modulo $\bQb$-exact terms, but because the generators $W_A$ have no terms linear in the $\phi_a$ all $\bQb$-exact terms have $\phi$-dependent coefficients.  In what follows, we work at the level of cohomology and leave any $\bQb$-exact terms implicit.

  It is then easy to repeat the argument that by a linear change of variables on the $\gamma^A$ we can take $R^A_B$ to be diagonal, so that
\begin{align}
\delta\gamma^A = \sigmab_A \gamma^A + \sum_{B} M^A_B(\phi) \gamma^B~,
\end{align}
and
\begin{align}
\gamma^{A,\sigmab_A} = \gamma^A + \sum_{B} G^A_B(\phi) \gamma^B~,
\end{align}
while $\gamma^{A,\qb_i}$ with $\qb_i \neq \sigmab_A$ have no $\phi$-independent terms.

As before we find charges $r_A$ so that $\gamma^A$ belongs to $\cF^{\,r_A}$ but not to $ \cF^{\qb}$ for $\qb > r_A$, and after a change of variables and reordering according to the $r_A$, we have
\begin{align}
\gamma^{A,\sigmab_A} = \gamma^A + \sum_{B<A} G^A_B(\phi)\gamma^B~.
\end{align}
Clearly the change of variables $\gamma^{'A} = \gamma^{A,\sigmab_A}$ is invertible and yields the desired eigenstates $\gamma^{'A}$ of the R-symmetry action, with $\delta \gamma^{'A} = \sigmab_A \gamma^{'A}$~.

\section{Koszul homology} \label{app:Koszul}
In this appendix we make some comments on unitarity in the B/2 ring of the (0,2) LG theory encoded by the Koszul homology associated to the ideal $\WW$.

Consider the sub-sector of $H$ consisting of fields constructed purely from the $\phi_a$.  This generates all fields in $H^0$ and has algebraic description as the quotient ring
\begin{align}
R_{\WW} = \C[\phi_1,\ldots,\phi_n] / \WW~.
\end{align}
For a zero-dimensional $\WW$ this is a finite-dimensional vector space, and when graded by the R-symmetry, it must have a field of top degree.  Assuming that we work with $\phi_a$ as eigenstates of $(J_{\tir},\Jb_{\tir})$, we expect that for $N=n$ the top charge should be $q_{\tir} = \qb_{\tir} = \cb /3$.  It is easy to write down an operator that saturates the bound and represents the unit spectral flow operator:
\begin{align}
\cO = \det_{a,b} W_{a,b}~,
\end{align} 
has charges $q=\qb = \cb/3$ and for any zero-dimensional ${\WW}$, $\cO \neq 0 \in R_{\WW}$~\cite{Tsikh:2004rc}.  Moreover, $R_{\WW}$ has dimension $1$ at top degree~\cite{MR2896292}.

More generally, for any $N \ge n$ we can consider operators of the form
\begin{align}
\cO[\omega] = \omega_{A_1\cdots A_k}(\phi) \gammab^{A_1} \cdots \gammab^{A_k}~.
\end{align}
Taking $\bQb$-cohomology, this set of operators leads to the Koszul homology $H_k(K_{\bullet},\WW)$ of the ideal $\WW$ that is isomorphic to the B/2 topological heterotic ring of the theory~\cite{Kawai:1994qy,Melnikov:2009nh,Melnikov:2019tpl}.  It follows that $H_0 (K_{\bullet},\WW) = R_{\WW}$.  When $\WW$ is zero-dimensional, commutative algebra results~\cite{Bruns:1993cm,Eisenbud:1995ca} show that
\begin{align}
H_k(K_{\bullet},\WW) = \begin{cases} 0 & \text{for}~~k> N-n \nonumber\\ \text{non-zero} & \text{for}~~ k= N-n \end{cases}~.
\end{align}
Since we have a non-chiral spectral flow even in theories with $N>n$, we expect that $H_{N-n}(K_{\bullet},\WW)$ is one-dimensional and related to spectral flow of the identity.  It is not hard to construct a representative:  set $k = N-n$ and take $\cO[\omega]$ with
\begin{align}
\omega_{A_1\cdots A_{k}} & = \sum_{B_1,\ldots,B_n} Q_{A_1}Q_{A_2} \cdots Q_{A_{k}} \ep^{A_1 \cdots A_k B_1 \cdots B_n} W_{B_{1},1} W_{B_{2},2} \cdots W_{B_{n},n}~,
\end{align}
where $\ep$ is the usual fully anti-symmetric rank $N$ tensor.  It is easy to show that $\bQb\cdot \cO[\omega] = 0$, and we expect this operator to generate $H_{N-n}(K_{\bullet},\WW)$, but we are not aware of a purely algebraic proof of this statement.

\section{Holomorphic symmetry currents for $N=n$}  \label{app:GenSym}
A holomorphic current $J$ may in principle have terms charged with respect to $\GU(1)_{\text{R},\tuv}$, in which case  the most general spin--$1$ candidate operator has the form
\begin{align}
J &=  \sum_{a=1}^n :A^a(\phi)\rho_a: +  \sum_{A,B =1}^N B_A^B(\phi) :\gamma^A\gammab_B:  \nonumber\\
   &\qquad + \sum_{a=1}^n C^{a}(\phi) \p\phi_a
                 + \sum_{A,B=1}^N  \left( D_{AB}(\phi)\gamma^A\gamma^B + E^{AB}\gammab_A\gammab_B \right)~.
\end{align}
We will now argue that the each term in the second line must be absent when $N=n$. 

Since the $\bQb$ action is graded by $\GU(1)_{\text{R,\tuv}}$, each term in the second line must be separately $\bQb$-closed.  The $C$--term must be zero by exactly the same argument as before:  it is inconsistent with $J$ having R-charge $0$.  As we discussed in appendix~\ref{app:Koszul}, the $E$--term must be $\bQb$-exact when $N=n$: otherwise the ideal $\WW$ cannot be zero dimensional.  

It remains to rule out the $D$--term.  As we showed in the last section of appendix~\ref{app:Rsym}, when $N=n$ and the ideal is zero-dimensional, there is a holomorphic field redefinition
\begin{align}
\gamma^{'A} = \sum_{B=1}^N F_A^B(\phi) \gamma^B
\end{align}
to fields $\gamma^{'A}$ that are R-symmetry eigenstates.  Writing the $D$--term in terms of the $\phi'_a$ and $\gamma^{'A}$ fields, and using the positivity of the R-charge for chiral primary states, we conclude that the $D$--term must be absent.

Thus, when $N=n$, every holomorphic symmetry $J$ must have a representative in $H^1$ of the form~(\ref{eq:genspin1}), and its closure is again equivalent to a holomorphic symmetry of the potential $\cW$  as in~(\ref{eq:Jclosed}).

\section{Explicitly Calculating Neutral Cohomology}\label{app:Tractors}
To check the possibility that the chiral algebra contains a spin-2 holomorphic current that might serve as $T_{\tir}$ and invalidate our conclusion that the chiral algebra is inconsistent with a compact, unitary SCFT determining the IR physics, we performed an explicit calculation of the cohomology groups $H^{s,\ast,0} = \oplus_{\qb_{\tuv}}H^{s, \qb_{\tuv},0}$ of $\bQb$ for $0\le s\le 2$  for the models of \ref{ss:badunicorn} with $p=5$ and $p=11$.  The calculation proceeded as follows:

\begin{enumerate}
\item All monomials in $\rho_a, \gamma^A, \gammab_A$ with spin $0\le s\le 5/2$ and non-positive $q$ were tabulated (monomials of degree higher than one in a given fermi field included suitable derivatives).
\item For each value of $S \in \{0,1/2,1,\dots, 5/2\}$, and for each of the monomials from step 1 with $s\le  S$ and $S-s\in\Z$, its products with all monomials in $\phi^a$ with charge $-q$ were tabulated, and all terms obtained from these by introducing $S-s$ holomorphic derivatives listed.  This produced the vector spaces $V^{S,\ast,0}$ of neutral monomials in fields and their derivatives with spin $S$ for $0\le S\le 5/2$ of dimensions $D_S = (1,5,15,48,131,347)$.
\item The action of $\bQb$ on each of these monomials was directly evaluated, allowing the determination of the matrix implementing the action $\bQb:V^{S,\ast,0}\to V^{S-1/2,\ast,0}$.  We find matrices of rank $r_S = (0,5,9,39,89)$.
\item These are consistent with dimensions $h_S = (1,0,1,0,3)$ for the cohomology groups in question.   To verify this we explicitly found a basis for
  \begin{align}
     \ker\left(\bQb\oplus\bQb^T: V^{S,\ast,0}\oplus V^{S,\ast,0}\to V^{S-1/2,\ast,0}\oplus V^{S+1/2,\ast,0}\right)\ .
  \end{align}
  We verified that the cohomology is concentrated at  $\qb_{\tuv}=0$ and is spanned by the expected currents $J$ at $S=1$, and by $T$, $\p J $, and $:JJ:$ at $S=2$.
  \end{enumerate}

\bibliographystyle{./utphys}
\bibliography{./newref}

\end{document}

%% file: basedefs2.tex
\DeclareFontFamily{U}{rsf}{}
\DeclareFontShape{U}{rsf}{m}{n}{
  <5> <6> rsfs5 <7> <8> <9> rsfs7 <10-> rsfs10}{}
\DeclareMathAlphabet\Scr{U}{rsf}{m}{n}

\def\CO#1#2{{[#1,#2]}}
\def\AC#1#2{{\{#1,#2\}}}

\def\iden{{\mathbbm 1}}

\def\cDb{{\overline{\cD}}}

\def\GUL{\GU(1)_{\text{L}}}

\def\C{{\mathbb C}}

\def\Z{{\mathbb Z}}

\def\SL{\operatorname{SL}}

\def\GU{\operatorname{U{}}}

\def\p{\partial}
\def\pb{\bar{\partial}}

\def\la{\langle}
\def\ra{\rangle}

\def\ff#1#2{{\textstyle\frac{#1}{#2}}}

\def\cA{{\cal A}}
\def\cB{{\cal B}}

\def\cD{{\cal D}}

\def\cF{{\cal F}}
\def\cG{{\cal G}}

\def\cI{{\cal I}}
\def\cJ{{\cal J}}
\def\cK{{\cal K}}

\def\cM{{\cal M}}
\def\cN{{\cal N}}
\def\cO{{\cal O}}

\def\cR{{\cal R}}

\def\cT{{\cal T}}
\def\cU{{\cal U}}
\def\cV{{\cal V}}
\def\cW{{\cal W}}

\def\ep{{\epsilon}}

\newcommand\deltah{\widehat{\delta}}

\newcommand\gammab{\overline{\gamma}}

\newcommand\thetab{\overline{\theta}}

\newcommand\lambdab{\overline{\lambda}}

\newcommand\sigmab{\overline{\sigma}}

\newcommand\phib{\overline{\phi}}

\newcommand\psib{\overline{\psi}}

\newcommand\Gammab{\overline{\Gamma}}

\newcommand\Phib{\overline{\Phi}}

\newcommand\cb{\overline{c}}

\newcommand\hb{\overline{h}}

\newcommand\qb{\overline{q}}

\newcommand\wb{\overline{w}}

\newcommand\zb{\overline{z}}

\newcommand\Mh{\widehat{M}}

\newcommand\Jb{\overline{J}}

\newcommand\Tb{\overline{T}}

\newcommand\Wb{\overline{W}}
